
\input fontch.tex

%
%
%
\def\unredoffs{} \def\redoffs{\voffset=-.31truein\hoffset=-.48truein}
\def\speclscape{}
%
%
%
%
%
\newbox\leftpage \newdimen\fullhsize \newdimen\hstitle \newdimen\hsbody
\tolerance=1000\hfuzz=2pt
\catcode`\@=11 
\ifx\hyperdef\UNd@FiNeD\def\hyperdef#1#2#3#4{#4}\def\hyperref#1#2#3#4{#4}\fi
\def\bigans{b }
\def\answ{b }
%
\ifx\answ\bigans\message{(This will come out unreduced.}
\magnification=1200\unredoffs\baselineskip=16pt plus 2pt minus 1pt
\hsbody=\hsize \hstitle=\hsize 
\else\message{(This will be reduced.} \let\l@r=L
\magnification=1000\baselineskip=16pt plus 2pt minus 1pt \vsize=7truein
\redoffs \hstitle=8truein\hsbody=4.75truein\fullhsize=10truein\hsize=\hsbody
\output={\ifnum\pageno=0 
  \shipout\vbox{\speclscape{\hsize\fullhsize\makeheadline}
    \hbox to \fullhsize{\hfill\pagebody\hfill}}\advancepageno
  \else
  \almostshipout{\leftline{\vbox{\pagebody\makefootline}}}\advancepageno
  \fi}
\def\almostshipout#1{\if L\l@r \count1=1 \message{[\the\count0.\the\count1]}
      \global\setbox\leftpage=#1 \global\let\l@r=R
 \else \count1=2
  \shipout\vbox{\speclscape{\hsize\fullhsize\makeheadline}
      \hbox to\fullhsize{\box\leftpage\hfil#1}}  \global\let\l@r=L\fi}
\fi
%
\newcount\yearltd\yearltd=\year\advance\yearltd by -2000

\def\Title#1#2{\nopagenumbers\abstractfont\hsize=\hstitle\rightline{#1}%
\vskip 1in\centerline{\titlefont #2}\abstractfont\vskip .5in\pageno=0}
\def\Date#1{\vfill\leftline{#1}\tenpoint\supereject\global\hsize=\hsbody%
\footline={\hss\tenrm\hyperdef\hypernoname{page}\folio\folio\hss}}%
%

\def\draftmode{\message{ DRAFTMODE }\def\draftdate{{\rm preliminary draft:
\number\month/\number\day/\number\yearltd\ \ \hourmin}}%
\headline={\hfil\draftdate}\writelabels\baselineskip=20pt plus 2pt minus 2pt
 {\count255=\time\divide\count255 by 60 \xdef\hourmin{\number\count255}
  \multiply\count255 by-60\advance\count255 by\time
  \xdef\hourmin{\hourmin:\ifnum\count255<10 0\fi\the\count255}}}
\def\nolabels{\def\wrlabeL##1{}\def\eqlabeL##1{}\def\reflabeL##1{}}
\def\writelabels{\def\wrlabeL##1{\leavevmode\vadjust{\rlap{\smash%
{\line{{\escapechar=` \hfill\rlap{\sevenrm\hskip.03in\string##1}}}}}}}%
\def\eqlabeL##1{{\escapechar-1\rlap{\sevenrm\hskip.05in\string##1}}}%
\def\reflabeL##1{\noexpand\llap{\noexpand\sevenrm\string\string\string##1}}}
\nolabels
%
\global\newcount\secno \global\secno=0
\global\newcount\meqno \global\meqno=1
\def\s@csym{}
\def\newsec#1{\global\advance\secno by1%
{\toks0{#1}\message{(\the\secno. \the\toks0)}}%
\global\subsecno=0\eqnres@t\let\s@csym\secsym\xdef\secn@m{\the\secno}\noindent
{\bf\hyperdef\hypernoname{section}{\the\secno}{\the\secno.} #1}%
\writetoca{{\string\hyperref{}{section}{\the\secno}{\the\secno.}} {#1}}%
\par\nobreak\medskip\nobreak}
\def\eqnres@t{\xdef\secsym{\the\secno.}\global\meqno=1\bigbreak\bigskip}
\def\sequentialequations{\def\eqnres@t{\bigbreak}}\xdef\secsym{}
\global\newcount\subsecno \global\subsecno=0
\def\subsec#1{\global\advance\subsecno by1%
{\toks0{#1}\message{(\s@csym\the\subsecno. \the\toks0)}}%
\ifnum\lastpenalty>9000\else\bigbreak\fi
\noindent{\it\hyperdef\hypernoname{subsection}{\secn@m.\the\subsecno}%
{\secn@m.\the\subsecno.} #1}\writetoca{\string\quad
{\string\hyperref{}{subsection}{\secn@m.\the\subsecno}{\secn@m.\the\subsecno.}}
{#1}}\par\nobreak\medskip\nobreak}
\def\appendix#1#2{\global\meqno=1\global\subsecno=0\xdef\secsym{\hbox{#1.}}%
\bigbreak\bigskip\noindent{\bf Appendix \hyperdef\hypernoname{appendix}{#1}%
{#1.} #2}{\toks0{(#1. #2)}\message{\the\toks0}}%
\xdef\s@csym{#1.}\xdef\secn@m{#1}%
\writetoca{\string\hyperref{}{appendix}{#1}{Appendix {#1.}} {#2}}%
\par\nobreak\medskip\nobreak}
%
%
\def\checkm@de#1#2{\ifmmode{\def\f@rst##1{##1}\hyperdef\hypernoname{equation}%
{#1}{#2}}\else\hyperref{}{equation}{#1}{#2}\fi}
\def\eqnn#1{\DefWarn#1\xdef #1{(\noexpand\relax\noexpand\checkm@de%
{\s@csym\the\meqno}{\secsym\the\meqno})}%
\wrlabeL#1\writedef{#1\leftbracket#1}\global\advance\meqno by1}
\def\f@rst#1{\c@t#1a\em@ark}\def\c@t#1#2\em@ark{#1}
\def\eqna#1{\DefWarn#1\wrlabeL{#1$\{\}$}%
\xdef #1##1{(\noexpand\relax\noexpand\checkm@de%
{\s@csym\the\meqno\noexpand\f@rst{##1}}{\hbox{$\secsym\the\meqno##1$}})}
\writedef{#1\numbersign1\leftbracket#1{\numbersign1}}\global\advance\meqno by1}
\def\eqn#1#2{\DefWarn#1%
\xdef #1{(\noexpand\hyperref{}{equation}{\s@csym\the\meqno}%
{\secsym\the\meqno})}$$#2\eqno(\hyperdef\hypernoname{equation}%
{\s@csym\the\meqno}{\secsym\the\meqno})\eqlabeL#1$$%
\writedef{#1\leftbracket#1}\global\advance\meqno by1}
\def\xeqn{\expandafter\xe@n}\def\xe@n(#1){#1}
\def\xeqna#1{\expandafter\xe@n#1}
\def\eqns#1{(\e@ns #1{\hbox{}})}
\def\e@ns#1{\ifx\UNd@FiNeD#1\message{eqnlabel \string#1 is undefined.}%
\xdef#1{(?.?)}\fi{\let\hyperref=\relax\xdef\next{#1}}%
\ifx\next\em@rk\def\next{}\else%
\ifx\next#1\xeqn#1\else\def\n@xt{#1}\ifx\n@xt\next#1\else\xeqna#1\fi
\fi\let\next=\e@ns\fi\next}

\def\DefWarn#1{\ifx\UNd@FiNeD#1\else
\immediate\write16{*** WARNING: the label \string#1 is already defined ***}\fi}
%
\newskip\footskip\footskip14pt plus 1pt minus 1pt 
\def\footnotefont{\ninepoint}\def\f@t#1{\footnotefont #1\@foot}
\def\f@@t{\baselineskip\footskip\bgroup\footnotefont\aftergroup\@foot\let\next}
\setbox\strutbox=\hbox{\vrule height9.5pt depth4.5pt width0pt}
\global\newcount\ftno \global\ftno=0
\def\foot{\global\advance\ftno by1\def\foot@rg{\hyperref{}{footnote}%
{\the\ftno}{\the\ftno}\xdef\foot@rg{\noexpand\hyperdef\noexpand\hypernoname%
{footnote}{\the\ftno}{\the\ftno}}}\footnote{$^{\foot@rg}$}}
%
\newwrite\ftfile
\def\footend{\def\foot{\global\advance\ftno by1\chardef\wfile=\ftfile
\hyperref{}{footnote}{\the\ftno}{$^{\the\ftno}$}%
\ifnum\ftno=1\immediate\openout\ftfile=\jobname.fts\fi%
\immediate\write\ftfile{\noexpand\smallskip%
\noexpand\item{\noexpand\hyperdef\noexpand\hypernoname{footnote}
{\the\ftno}{f\the\ftno}:\ }\pctsign}\findarg}%
\def\footatend{\vfill\eject\immediate\closeout\ftfile{\parindent=20pt
\centerline{\bf Footnotes}\nobreak\bigskip\input \jobname.fts }}}
\def\footatend{}
%
%
\global\newcount\refno \global\refno=1
\newwrite\rfile
\def\ref{[\hyperref{}{reference}{\the\refno}{\the\refno}]\nref}
\def\nref#1{\DefWarn#1%
\xdef#1{[\noexpand\hyperref{}{reference}{\the\refno}{\the\refno}]}%
\writedef{#1\leftbracket#1}%
\ifnum\refno=1\immediate\openout\rfile=\jobname.refs\fi
\chardef\wfile=\rfile\immediate\write\rfile{\noexpand\item{[\noexpand\hyperdef%
\noexpand\hypernoname{reference}{\the\refno}{\the\refno}]\ }%
\reflabeL{#1\hskip.31in}\pctsign}\global\advance\refno by1\findarg}
\def\findarg#1#{\begingroup\obeylines\newlinechar=`\^^M\pass@rg}
{\obeylines\gdef\pass@rg#1{\writ@line\relax #1^^M\hbox{}^^M}%
\gdef\writ@line#1^^M{\expandafter\toks0\expandafter{\striprel@x #1}%
\edef\next{\the\toks0}\ifx\next\em@rk\let\next=\endgroup\else\ifx\next\empty%
\else\immediate\write\wfile{\the\toks0}\fi\let\next=\writ@line\fi\next\relax}}
\def\striprel@x#1{} \def\em@rk{\hbox{}}
\def\lref{\begingroup\obeylines\lr@f}
\def\lr@f#1#2{\DefWarn#1\gdef#1{\let#1=\UNd@FiNeD\ref#1{#2}}\endgroup\unskip}

\def\addref#1{\immediate\write\rfile{\noexpand\item{}#1}} 
\def\listrefs{\footatend\vfill\supereject\immediate\closeout\rfile\writestoppt
\baselineskip=\footskip\centerline{{\bf References}}\bigskip{\parindent=20pt%
\frenchspacing\escapechar=` \input \jobname.refs\vfill\eject}\nonfrenchspacing}
\def\startrefs#1{\immediate\openout\rfile=\jobname.refs\refno=#1}
\def\xref{\expandafter\xr@f}\def\xr@f[#1]{#1}
\def\refs#1{\count255=1[\r@fs #1{\hbox{}}]}
\def\r@fs#1{\ifx\UNd@FiNeD#1\message{reflabel \string#1 is undefined.}%
\nref#1{need to supply reference \string#1.}\fi%
\vphantom{\hphantom{#1}}{\let\hyperref=\relax\xdef\next{#1}}%
\ifx\next\em@rk\def\next{}%
\else\ifx\next#1\ifodd\count255\relax\xref#1\count255=0\fi%
\else#1\count255=1\fi\let\next=\r@fs\fi\next}
%

%
\newwrite\ffile\global\newcount\figno \global\figno=1
\def\fig{fig.~\hyperref{}{figure}{\the\figno}{\the\figno}\nfig}
\def\nfig#1{\DefWarn#1%
\xdef#1{fig.~\noexpand\hyperref{}{figure}{\the\figno}{\the\figno}}%
\writedef{#1\leftbracket fig.\noexpand~\xfig#1}%
\ifnum\figno=1\immediate\openout\ffile=\jobname.figs\fi\chardef\wfile=\ffile%
{\let\hyperref=\relax
\immediate\write\ffile{\noexpand\medskip\noexpand\item{Fig.\ %
\noexpand\hyperdef\noexpand\hypernoname{figure}{\the\figno}{\the\figno}. }
\reflabeL{#1\hskip.55in}\pctsign}}\global\advance\figno by1\findarg}
\def\listfigs{\vfill\eject\immediate\closeout\ffile{\parindent40pt
\baselineskip14pt\centerline{{\bf Figure Captions}}\nobreak\medskip
\escapechar=` \input \jobname.figs\vfill\eject}}
\def\xfig{\expandafter\xf@g}\def\xf@g fig.\penalty\@M\ {}
\def\figs#1{figs.~\f@gs #1{\hbox{}}}
\def\f@gs#1{{\let\hyperref=\relax\xdef\next{#1}}\ifx\next\em@rk\def\next{}\else
\ifx\next#1\xfig #1\else#1\fi\let\next=\f@gs\fi\next}
\def\figin{\epsfcheck\figin}\def\figins{\epsfcheck\figins}
\def\epsfcheck{\ifx\epsfbox\UNd@FiNeD
\message{(NO epsf.tex, FIGURES WILL BE IGNORED)}
\gdef\figin##1{\vskip2in}\gdef\figins##1{\hskip.5in}
\else\message{(FIGURES WILL BE INCLUDED)}%
\gdef\figin##1{##1}\gdef\figins##1{##1}\fi}
\def\DefWarn#1{}
\def\figinsert{\goodbreak\midinsert}
\def\ifig#1#2#3{\DefWarn#1\xdef#1{fig.~\noexpand\hyperref{}{figure}%
{\the\figno}{\the\figno}}\writedef{#1\leftbracket fig.\noexpand~\xfig#1}%
\figinsert\figin{\centerline{#3}}\medskip\centerline{\vbox{\baselineskip12pt
\advance\hsize by -1truein\noindent\wrlabeL{#1=#1}\footnotefont%
{\bf Fig.~\hyperdef\hypernoname{figure}{\the\figno}{\the\figno}:} #2}}
\bigskip\endinsert\global\advance\figno by1}
\newwrite\lfile
{\escapechar-1\xdef\pctsign{\string\%}\xdef\leftbracket{\string\{}
\xdef\rightbracket{\string\}}\xdef\numbersign{\string\#}}
\def\writedefs{\immediate\openout\lfile=\jobname.defs \def\writedef##1{%
{\let\hyperref=\relax\let\hyperdef=\relax\let\hypernoname=\relax
 \immediate\write\lfile{\string\def\string##1\rightbracket}}}}%
\def\writestop{\def\writestoppt{\immediate\write\lfile{\string\pageno
 \the\pageno\string\startrefs\leftbracket\the\refno\rightbracket
 \string\def\string\secsym\leftbracket\secsym\rightbracket
 \string\secno\the\secno\string\meqno\the\meqno}\immediate\closeout\lfile}}
\def\writestoppt{}\def\writedef#1{}
\def\seclab#1{\DefWarn#1%
\xdef #1{\noexpand\hyperref{}{section}{\the\secno}{\the\secno}}%
\writedef{#1\leftbracket#1}\wrlabeL{#1=#1}}
\def\subseclab#1{\DefWarn#1%
\xdef #1{\noexpand\hyperref{}{subsection}{\secn@m.\the\subsecno}%
{\secn@m.\the\subsecno}}\writedef{#1\leftbracket#1}\wrlabeL{#1=#1}}
\def\applab#1{\DefWarn#1%
\xdef #1{\noexpand\hyperref{}{appendix}{\secn@m}{\secn@m}}%
\writedef{#1\leftbracket#1}\wrlabeL{#1=#1}}
\newwrite\tfile \def\writetoca#1{}
\def\leaderfill{\leaders\hbox to 1em{\hss.\hss}\hfill}
\def\writetoc{\immediate\openout\tfile=\jobname.toc
   \def\writetoca##1{{\edef\next{\write\tfile{\noindent ##1
   \string\leaderfill {\string\hyperref{}{page}{\noexpand\number\pageno}%
                       {\noexpand\number\pageno}} \par}}\next}}}
\newread\ch@ckfile
\def\listtoc{\immediate\closeout\tfile\immediate\openin\ch@ckfile=\jobname.toc
\ifeof\ch@ckfile\message{no file \jobname.toc, no table of contents this pass}%
\else\closein\ch@ckfile\centerline{\bf Contents}\nobreak\medskip%
{\baselineskip=12pt\footnotefont\parskip=0pt\catcode`\@=11\input\jobname.toc
\catcode`\@=12\bigbreak\bigskip}\fi}
\catcode`\@=12 
%
\edef\tfontsize{\ifx\answ\bigans scaled\magstep3\else scaled\magstep4\fi}
\font\titlerm=cmr10 \tfontsize \font\titlerms=cmr7 \tfontsize
\font\titlermss=cmr5 \tfontsize \font\titlei=cmmi10 \tfontsize
\font\titleis=cmmi7 \tfontsize \font\titleiss=cmmi5 \tfontsize
\font\titlesy=cmsy10 \tfontsize \font\titlesys=cmsy7 \tfontsize
\font\titlesyss=cmsy5 \tfontsize \font\titleit=cmti10 \tfontsize
\skewchar\titlei='177 \skewchar\titleis='177 \skewchar\titleiss='177
\skewchar\titlesy='60 \skewchar\titlesys='60 \skewchar\titlesyss='60
\def\titlefont{\def\rm{\fam0\titlerm}
\textfont0=\titlerm \scriptfont0=\titlerms \scriptscriptfont0=\titlermss
\textfont1=\titlei \scriptfont1=\titleis \scriptscriptfont1=\titleiss
\textfont2=\titlesy \scriptfont2=\titlesys \scriptscriptfont2=\titlesyss
\textfont\itfam=\titleit \def\it{\fam\itfam\titleit}\rm}
 \ifx\answ\bigans\else scaled\magstep1\fi
\ifx\answ\bigans\def\abstractfont{\tenpoint}\else
\font\absit=cmti10 scaled \magstep1
\font\abssl=cmsl10 scaled \magstep1
\font\absrm=cmr10 scaled\magstep1 \font\absrms=cmr7 scaled\magstep1
\font\absrmss=cmr5 scaled\magstep1 \font\absi=cmmi10 scaled\magstep1
\font\absis=cmmi7 scaled\magstep1 \font\absiss=cmmi5 scaled\magstep1
\font\abssy=cmsy10 scaled\magstep1 \font\abssys=cmsy7 scaled\magstep1
\font\abssyss=cmsy5 scaled\magstep1 \font\absbf=cmbx10 scaled\magstep1
\skewchar\absi='177 \skewchar\absis='177 \skewchar\absiss='177
\skewchar\abssy='60 \skewchar\abssys='60 \skewchar\abssyss='60
\def\abstractfont{\def\rm{\fam0\absrm}
\textfont0=\absrm \scriptfont0=\absrms \scriptscriptfont0=\absrmss
\textfont1=\absi \scriptfont1=\absis \scriptscriptfont1=\absiss
\textfont2=\abssy \scriptfont2=\abssys \scriptscriptfont2=\abssyss
\textfont\itfam=\absit \def\it{\fam\itfam\absit}\def\footnotefont{\tenpoint}%
\textfont\slfam=\abssl \def\sl{\fam\slfam\abssl}%
\textfont\bffam=\absbf \def\bf{\fam\bffam\absbf}\rm}\fi
\def\tenpoint{\def\rm{\fam0\tenrm}
\textfont0=\tenrm \scriptfont0=\sevenrm \scriptscriptfont0=\fiverm
\textfont1=\teni  \scriptfont1=\seveni  \scriptscriptfont1=\fivei
\textfont2=\tensy \scriptfont2=\sevensy \scriptscriptfont2=\fivesy
\textfont\itfam=\tenit \def\it{\fam\itfam\tenit}\def\footnotefont{\ninepoint}%
\textfont\bffam=\tenbf \def\bf{\fam\bffam\tenbf}\def\sl{\fam\slfam\tensl}\rm}
\font\ninerm=cmr9 \font\sixrm=cmr6 \font\ninei=cmmi9 \font\sixi=cmmi6
\font\ninesy=cmsy9 \font\sixsy=cmsy6 \font\ninebf=cmbx9
\font\nineit=cmti9 \font\ninesl=cmsl9 \skewchar\ninei='177
\skewchar\sixi='177 \skewchar\ninesy='60 \skewchar\sixsy='60
\def\ninepoint{\def\rm{\fam0\ninerm}
\textfont0=\ninerm \scriptfont0=\sixrm \scriptscriptfont0=\fiverm
\textfont1=\ninei \scriptfont1=\sixi \scriptscriptfont1=\fivei
\textfont2=\ninesy \scriptfont2=\sixsy \scriptscriptfont2=\fivesy
\textfont\itfam=\ninei \def\it{\fam\itfam\nineit}\def\sl{\fam\slfam\ninesl}%
\textfont\bffam=\ninebf \def\bf{\fam\bffam\ninebf}\rm}
%
%

\hyphenation{anom-aly anom-alies coun-ter-term coun-ter-terms}
\def\inv{^{\raise.15ex\hbox{${\scriptscriptstyle -}$}\kern-.05em 1}}

\def\Dsl{\,\raise.15ex\hbox{/}\mkern-13.5mu D} 
\def\dsl{\raise.15ex\hbox{/}\kern-.57em\partial}

 \def\Tr{{\rm Tr}}
\def\lspace{\ifx\answ\bigans{}\else\qquad\fi}
\def\lbspace{\ifx\answ\bigans{}\else\hskip-.2in\fi} 
\def\boxeqn#1{\vcenter{\vbox{\hrule\hbox{\vrule\kern3pt\vbox{\kern3pt
	\hbox{${\displaystyle #1}$}\kern3pt}\kern3pt\vrule}\hrule}}}
\def\mbox#1#2{\vcenter{\hrule \hbox{\vrule height#2in
		\kern#1in \vrule} \hrule}}  
%

\def\darr#1{\raise1.5ex\hbox{$\leftrightarrow$}\mkern-16.5mu #1}

\def\roughly#1{\raise.3ex\hbox{$#1$\kern-.75em\lower1ex\hbox{$\sim$}}}

\def\bb{
\font\tenmsb=msbm10
\font\sevenmsb=msbm7
\font\fivemsb=msbm5
\textfont1=\tenmsb
\scriptfont1=\sevenmsb
\scriptscriptfont1=\fivemsb
}

\input amssym

\input epsf

\def\IZ{\relax\ifmmode\mathchoice
{\hbox{\cmss Z\kern-.4em Z}}{\hbox{\cmss Z\kern-.4em Z}} {\lower.9pt\hbox{\cmsss Z\kern-.4em Z}}
{\lower1.2pt\hbox{\cmsss Z\kern-.4em Z}}\else{\cmss Z\kern-.4em Z}\fi}

\newif\ifdraft\draftfalse
\newif\ifinter\interfalse
\ifdraft\draftmode\else\interfalse\fi
\def\journal#1&#2(#3){\unskip, \sl #1\ \bf #2 \rm(19#3) }
\def\andjournal#1&#2(#3){\sl #1~\bf #2 \rm (19#3) }

\def\frac#1#2{{#1\over#2}}

\def\inbar{\,\vrule height1.5ex width.4pt depth0pt}
\def\IC{\relax\hbox{$\inbar\kern-.3em{\rm C}$}}
\def\IR{\relax{\rm I\kern-.18em R}}
\def\IP{\relax{\rm I\kern-.18em P}}

%
%


%
\catcode`\@=11
\def\slash#1{\mathord{\mathpalette\c@ncel{#1}}}
\overfullrule=0pt

\def\underrel#1\over#2{\mathrel{\mathop{\kern\z@#1}\limits_{#2}}}

\catcode`\@=12


%


\def\[{[}
\def\]{]}

\def\comment#1{ }

%
\def\draftnote#1{\ifdraft{\baselineskip2ex
                 \vbox{\kern1em\hrule\hbox{\vrule\kern1em\vbox{\kern1ex
                 \noindent \underbar{NOTE}: #1
             \vskip1ex}\kern1em\vrule}\hrule}}\fi}
\def\internote#1{\ifinter{\baselineskip2ex
                 \vbox{\kern1em\hrule\hbox{\vrule\kern1em\vbox{\kern1ex
                 \noindent \underbar{Internal Note}: #1
             \vskip1ex}\kern1em\vrule}\hrule}}\fi}

%
%



%
%
%
%

%

\def\inv{^{-1}}


\def\Tr{{\rm Tr}}

\def\1{{\ds 1}}

\def\S{\hbox{$\bb S$}}

\newfam\frakfam
\font\teneufm=eufm10
\font\seveneufm=eufm7
\font\fiveeufm=eufm5
\textfont\frakfam=\teneufm
\scriptfont\frakfam=\seveneufm
\scriptscriptfont\frakfam=\fiveeufm
\def\frak{\fam\frakfam \teneufm}

\lref\NiarchosAH{
  V.~Niarchos,
  ``Seiberg dualities and the 3d/4d connection,''
JHEP {\bf 1207}, 075 (2012).
[arXiv:1205.2086 [hep-th]].
}

\lref\AharonyGP{
  O.~Aharony,
  ``IR duality in d = 3 N=2 supersymmetric USp(2N(c)) and U(N(c)) gauge theories,''
Phys.\ Lett.\ B {\bf 404}, 71 (1997).
[hep-th/9703215].
}

\lref\AffleckAS{
  I.~Affleck, J.~A.~Harvey and E.~Witten,
  ``Instantons and (Super)Symmetry Breaking in (2+1)-Dimensions,''
Nucl.\ Phys.\ B {\bf 206}, 413 (1982)..
}

\lref\IntriligatorID{
  K.~A.~Intriligator and N.~Seiberg,
  ``Duality, monopoles, dyons, confinement and oblique confinement in supersymmetric SO(N(c)) gauge theories,''
Nucl.\ Phys.\ B {\bf 444}, 125 (1995).
[hep-th/9503179].
}

\lref\PasquettiFJ{
  S.~Pasquetti,
  ``Factorisation of N = 2 Theories on the Squashed 3-Sphere,''
JHEP {\bf 1204}, 120 (2012).
[arXiv:1111.6905 [hep-th]].
}

\lref\BeemMB{
  C.~Beem, T.~Dimofte and S.~Pasquetti,
  ``Holomorphic Blocks in Three Dimensions,''
[arXiv:1211.1986 [hep-th]].
}

\lref\SeibergPQ{
  N.~Seiberg,
  ``Electric - magnetic duality in supersymmetric nonAbelian gauge theories,''
Nucl.\ Phys.\ B {\bf 435}, 129 (1995).
[hep-th/9411149].
}

\lref\AharonyBX{
  O.~Aharony, A.~Hanany, K.~A.~Intriligator, N.~Seiberg and M.~J.~Strassler,
  ``Aspects of N=2 supersymmetric gauge theories in three-dimensions,''
Nucl.\ Phys.\ B {\bf 499}, 67 (1997).
[hep-th/9703110].
}

\lref\IntriligatorNE{
  K.~A.~Intriligator and P.~Pouliot,
  ``Exact superpotentials, quantum vacua and duality in supersymmetric SP(N(c)) gauge theories,''
Phys.\ Lett.\ B {\bf 353}, 471 (1995).
[hep-th/9505006].
}

\lref\KarchUX{
  A.~Karch,
  ``Seiberg duality in three-dimensions,''
Phys.\ Lett.\ B {\bf 405}, 79 (1997).
[hep-th/9703172].
}

\lref\SafdiRE{
  B.~R.~Safdi, I.~R.~Klebanov and J.~Lee,
  ``A Crack in the Conformal Window,''
[arXiv:1212.4502 [hep-th]].
}

\lref\SchweigertTG{
  C.~Schweigert,
  ``On moduli spaces of flat connections with nonsimply connected structure group,''
Nucl.\ Phys.\ B {\bf 492}, 743 (1997).
[hep-th/9611092].
}

\lref\GiveonZN{
  A.~Giveon and D.~Kutasov,
  ``Seiberg Duality in Chern-Simons Theory,''
Nucl.\ Phys.\ B {\bf 812}, 1 (2009).
[arXiv:0808.0360 [hep-th]].
}

\lref\GaiottoBE{
  D.~Gaiotto, G.~W.~Moore and A.~Neitzke,
  ``Framed BPS States,''
[arXiv:1006.0146 [hep-th]].
}

\lref\AldayRS{
  L.~F.~Alday, M.~Bullimore and M.~Fluder,
  ``On S-duality of the Superconformal Index on Lens Spaces and 2d TQFT,''
JHEP {\bf 1305}, 122 (2013).
[arXiv:1301.7486 [hep-th]].
}

\lref\RazamatJXA{
  S.~S.~Razamat and M.~Yamazaki,
  ``S-duality and the N=2 Lens Space Index,''
[arXiv:1306.1543 [hep-th]].
}

\lref\NiarchosAH{
  V.~Niarchos,
  ``Seiberg dualities and the 3d/4d connection,''
JHEP {\bf 1207}, 075 (2012).
[arXiv:1205.2086 [hep-th]].
}

\lref\almost{
  A.~Borel, R.~Friedman, J.~W.~Morgan,
  ``Almost commuting elements in compact Lie groups,''
arXiv:math/9907007.
}

\lref\KapustinJM{
  A.~Kapustin and B.~Willett,
  ``Generalized Superconformal Index for Three Dimensional Field Theories,''
[arXiv:1106.2484 [hep-th]].
}

\lref\AharonyGP{
  O.~Aharony,
  ``IR duality in d = 3 N=2 supersymmetric USp(2N(c)) and U(N(c)) gauge theories,''
Phys.\ Lett.\ B {\bf 404}, 71 (1997).
[hep-th/9703215].
}

\lref\FestucciaWS{
  G.~Festuccia and N.~Seiberg,
  ``Rigid Supersymmetric Theories in Curved Superspace,''
JHEP {\bf 1106}, 114 (2011).
[arXiv:1105.0689 [hep-th]].
}

\lref\RomelsbergerEG{
  C.~Romelsberger,
  ``Counting chiral primaries in N = 1, d=4 superconformal field theories,''
Nucl.\ Phys.\ B {\bf 747}, 329 (2006).
[hep-th/0510060].
}

\lref\KapustinKZ{
  A.~Kapustin, B.~Willett and I.~Yaakov,
  ``Exact Results for Wilson Loops in Superconformal Chern-Simons Theories with Matter,''
JHEP {\bf 1003}, 089 (2010).
[arXiv:0909.4559 [hep-th]].
}

\lref\DolanQI{
  F.~A.~Dolan and H.~Osborn,
  ``Applications of the Superconformal Index for Protected Operators and q-Hypergeometric Identities to N=1 Dual Theories,''
Nucl.\ Phys.\ B {\bf 818}, 137 (2009).
[arXiv:0801.4947 [hep-th]].
}

\lref\GaddeIA{
  A.~Gadde and W.~Yan,
  ``Reducing the 4d Index to the $S^3$ Partition Function,''
JHEP {\bf 1212}, 003 (2012).
[arXiv:1104.2592 [hep-th]].
}

\lref\DolanRP{
  F.~A.~H.~Dolan, V.~P.~Spiridonov and G.~S.~Vartanov,
  ``From 4d superconformal indices to 3d partition functions,''
Phys.\ Lett.\ B {\bf 704}, 234 (2011).
[arXiv:1104.1787 [hep-th]].
}

\lref\ImamuraUW{
  Y.~Imamura,
 ``Relation between the 4d superconformal index and the $S^3$ partition function,''
JHEP {\bf 1109}, 133 (2011).
[arXiv:1104.4482 [hep-th]].
}

\lref\HamaEA{
  N.~Hama, K.~Hosomichi and S.~Lee,
  ``SUSY Gauge Theories on Squashed Three-Spheres,''
JHEP {\bf 1105}, 014 (2011).
[arXiv:1102.4716 [hep-th]].
}

\lref\GaddeEN{
  A.~Gadde, L.~Rastelli, S.~S.~Razamat and W.~Yan,
  ``On the Superconformal Index of N=1 IR Fixed Points: A Holographic Check,''
JHEP {\bf 1103}, 041 (2011).
[arXiv:1011.5278 [hep-th]].
}

\lref\EagerHX{
  R.~Eager, J.~Schmude and Y.~Tachikawa,
  ``Superconformal Indices, Sasaki-Einstein Manifolds, and Cyclic Homologies,''
[arXiv:1207.0573 [hep-th]].
}

\lref\AffleckAS{
  I.~Affleck, J.~A.~Harvey and E.~Witten,
  ``Instantons and (Super)Symmetry Breaking in (2+1)-Dimensions,''
Nucl.\ Phys.\ B {\bf 206}, 413 (1982)..
}

\lref\SeibergPQ{
  N.~Seiberg,
  ``Electric - magnetic duality in supersymmetric nonAbelian gauge theories,''
Nucl.\ Phys.\ B {\bf 435}, 129 (1995).
[hep-th/9411149].
}

\lref\debult{
  F.~van~de~Bult,
  ``Hyperbolic Hypergeometric Functions,''
University of Amsterdam Ph.D. thesis
}

\lref\Shamirthesis{
  I.~Shamir,
  ``Aspects of three dimensional Seiberg duality,''
  M. Sc. thesis submitted to the Weizmann Institute of Science, April 2010.
  }

\lref\slthreeZ{
  J.~Felder, A.~Varchenko,
  ``The elliptic gamma function and $SL(3,Z) \times Z^3$,'' $\;\;$
[arXiv:math/0001184].
}

\lref\BeniniNC{
  F.~Benini, T.~Nishioka and M.~Yamazaki,
  ``4d Index to 3d Index and 2d TQFT,''
Phys.\ Rev.\ D {\bf 86}, 065015 (2012).
[arXiv:1109.0283 [hep-th]].
}

\lref\GaiottoWE{
  D.~Gaiotto,
  ``N=2 dualities,''
  JHEP {\bf 1208}, 034 (2012).
  [arXiv:0904.2715 [hep-th]].
}

\lref\SpiridonovZA{
  V.~P.~Spiridonov and G.~S.~Vartanov,
  ``Elliptic Hypergeometry of Supersymmetric Dualities,''
Commun.\ Math.\ Phys.\  {\bf 304}, 797 (2011).
[arXiv:0910.5944 [hep-th]].
}

\lref\BeniniMF{
  F.~Benini, C.~Closset and S.~Cremonesi,
  ``Comments on 3d Seiberg-like dualities,''
JHEP {\bf 1110}, 075 (2011).
[arXiv:1108.5373 [hep-th]].
}

\lref\ClossetVP{
  C.~Closset, T.~T.~Dumitrescu, G.~Festuccia, Z.~Komargodski and N.~Seiberg,
  ``Comments on Chern-Simons Contact Terms in Three Dimensions,''
JHEP {\bf 1209}, 091 (2012).
[arXiv:1206.5218 [hep-th]].
}

\lref\SpiridonovHF{
  V.~P.~Spiridonov and G.~S.~Vartanov,
  ``Elliptic hypergeometry of supersymmetric dualities II. Orthogonal groups, knots, and vortices,''
[arXiv:1107.5788 [hep-th]].
}

\lref\SpiridonovWW{
  V.~P.~Spiridonov and G.~S.~Vartanov,
  ``Elliptic hypergeometric integrals and 't Hooft anomaly matching conditions,''
JHEP {\bf 1206}, 016 (2012).
[arXiv:1203.5677 [hep-th]].
}

\lref\DimoftePY{
  T.~Dimofte, D.~Gaiotto and S.~Gukov,
  ``3-Manifolds and 3d Indices,''
[arXiv:1112.5179 [hep-th]].
}

\lref\KimWB{
  S.~Kim,
  ``The Complete superconformal index for N=6 Chern-Simons theory,''
Nucl.\ Phys.\ B {\bf 821}, 241 (2009), [Erratum-ibid.\ B {\bf 864}, 884 (2012)].
[arXiv:0903.4172 [hep-th]].
}

\lref\WillettGP{
  B.~Willett and I.~Yaakov,
  ``N=2 Dualities and Z Extremization in Three Dimensions,''
[arXiv:1104.0487 [hep-th]].
}

\lref\ImamuraSU{
  Y.~Imamura and S.~Yokoyama,
  ``Index for three dimensional superconformal field theories with general R-charge assignments,''
JHEP {\bf 1104}, 007 (2011).
[arXiv:1101.0557 [hep-th]].
}

\lref\FreedYA{
  D.~S.~Freed, G.~W.~Moore and G.~Segal,
  ``The Uncertainty of Fluxes,''
Commun.\ Math.\ Phys.\  {\bf 271}, 247 (2007).
[hep-th/0605198].
}

\lref\HwangQT{
  C.~Hwang, H.~Kim, K.~-J.~Park and J.~Park,
  ``Index computation for 3d Chern-Simons matter theory: test of Seiberg-like duality,''
JHEP {\bf 1109}, 037 (2011).
[arXiv:1107.4942 [hep-th]].
}

\lref\GreenDA{
  D.~Green, Z.~Komargodski, N.~Seiberg, Y.~Tachikawa and B.~Wecht,
  ``Exactly Marginal Deformations and Global Symmetries,''
JHEP {\bf 1006}, 106 (2010).
[arXiv:1005.3546 [hep-th]].
}

\lref\GaiottoXA{
  D.~Gaiotto, L.~Rastelli and S.~S.~Razamat,
  ``Bootstrapping the superconformal index with surface defects,''
[arXiv:1207.3577 [hep-th]].
}

\lref\IntriligatorID{
  K.~A.~Intriligator and N.~Seiberg,
  ``Duality, monopoles, dyons, confinement and oblique confinement in supersymmetric SO(N(c)) gauge theories,''
Nucl.\ Phys.\ B {\bf 444}, 125 (1995).
[hep-th/9503179].
}

\lref\SeibergNZ{
  N.~Seiberg and E.~Witten,
  ``Gauge dynamics and compactification to three-dimensions,''
In *Saclay 1996, The mathematical beauty of physics* 333-366.
[hep-th/9607163].
}

\lref\KinneyEJ{
  J.~Kinney, J.~M.~Maldacena, S.~Minwalla and S.~Raju,
  ``An Index for 4 dimensional super conformal theories,''
  Commun.\ Math.\ Phys.\  {\bf 275}, 209 (2007).
  [hep-th/0510251].
}

\lref\NakayamaUR{
  Y.~Nakayama,
  ``Index for supergravity on AdS(5) x T**1,1 and conifold gauge theory,''
Nucl.\ Phys.\ B {\bf 755}, 295 (2006).
[hep-th/0602284].
}

\lref\GaddeKB{
  A.~Gadde, E.~Pomoni, L.~Rastelli and S.~S.~Razamat,
  ``S-duality and 2d Topological QFT,''
JHEP {\bf 1003}, 032 (2010).
[arXiv:0910.2225 [hep-th]].
}

\lref\GaddeTE{
  A.~Gadde, L.~Rastelli, S.~S.~Razamat and W.~Yan,
  ``The Superconformal Index of the $E_6$ SCFT,''
JHEP {\bf 1008}, 107 (2010).
[arXiv:1003.4244 [hep-th]].
}

\lref\AharonyCI{
  O.~Aharony and I.~Shamir,
  ``On $O(N_c)$ d=3 N=2 supersymmetric QCD Theories,''
JHEP {\bf 1112}, 043 (2011).
[arXiv:1109.5081 [hep-th]].
}

\lref\GiveonSR{
  A.~Giveon and D.~Kutasov,
  ``Brane dynamics and gauge theory,''
Rev.\ Mod.\ Phys.\  {\bf 71}, 983 (1999).
[hep-th/9802067].
}

\lref\SpiridonovQV{
  V.~P.~Spiridonov and G.~S.~Vartanov,
  ``Superconformal indices of ${\cal N}=4$ SYM field theories,''
Lett.\ Math.\ Phys.\  {\bf 100}, 97 (2012).
[arXiv:1005.4196 [hep-th]].
}
\lref\GaddeUV{
  A.~Gadde, L.~Rastelli, S.~S.~Razamat and W.~Yan,
  ``Gauge Theories and Macdonald Polynomials,''
Commun.\ Math.\ Phys.\  {\bf 319}, 147 (2013).
[arXiv:1110.3740 [hep-th]].
}
\lref\KapustinGH{
  A.~Kapustin,
  ``Seiberg-like duality in three dimensions for orthogonal gauge groups,''
[arXiv:1104.0466 [hep-th]].
}

\lref\orthogpaper{O. Aharony, S. S. Razamat, N.~Seiberg and B.~Willett, 
``3d dualities from 4d dualities for orthogonal groups,''
[arXiv:1307.0511 [hep-th]].
}

\lref\readinglines{
  O.~Aharony, N.~Seiberg and Y.~Tachikawa,
  ``Reading between the lines of four-dimensional gauge theories,''
[arXiv:1305.0318 [hep-th]].
}

\lref\WittenNV{
  E.~Witten,
  ``Supersymmetric index in four-dimensional gauge theories,''
Adv.\ Theor.\ Math.\ Phys.\  {\bf 5}, 841 (2002).
[hep-th/0006010].
}

\lref\GaddeUV{
  A.~Gadde, L.~Rastelli, S.~S.~Razamat and W.~Yan,
  ``Gauge Theories and Macdonald Polynomials,''
Commun.\ Math.\ Phys.\  {\bf 319}, 147 (2013).
[arXiv:1110.3740 [hep-th]].
}

\lref\GaddeIK{
  A.~Gadde, L.~Rastelli, S.~S.~Razamat and W.~Yan,
  ``The 4d Superconformal Index from q-deformed 2d Yang-Mills,''
Phys.\ Rev.\ Lett.\  {\bf 106}, 241602 (2011).
[arXiv:1104.3850 [hep-th]].
}

\lref\GaiottoXA{
  D.~Gaiotto, L.~Rastelli and S.~S.~Razamat,
  ``Bootstrapping the superconformal index with surface defects,''
JHEP {\bf 1301}, 022 (2013).
[arXiv:1207.3577 [hep-th]].
}

\lref\GaiottoUQ{
  D.~Gaiotto and S.~S.~Razamat,
  ``Exceptional Indices,''
JHEP {\bf 1205}, 145 (2012).
[arXiv:1203.5517 [hep-th]].
}

\lref\RazamatUV{
  S.~S.~Razamat,
  ``On a modular property of N=2 superconformal theories in four dimensions,''
JHEP {\bf 1210}, 191 (2012).
[arXiv:1208.5056 [hep-th]].
}

\lref\noumi{
  Y.~Komori, M.~Noumi, J.~Shiraishi,
  ``Kernel Functions for Difference Operators of Ruijsenaars Type and Their Applications,''
SIGMA 5 (2009), 054.
[arXiv:0812.0279 [math.QA]].
}

\lref\RazamatJXA{
  S.~S.~Razamat and M.~Yamazaki,
  ``S-duality and the N=2 Lens Space Index,''
[arXiv:1306.1543 [hep-th]].
}

\lref\RazamatOPA{
  S.~S.~Razamat and B.~Willett,
  ``Global Properties of Supersymmetric Theories and the Lens Space,''
[arXiv:1307.4381 [hep-th]].
}

\lref\GaddeTE{
  A.~Gadde, L.~Rastelli, S.~S.~Razamat and W.~Yan,
  ``The Superconformal Index of the $E_6$ SCFT,''
JHEP {\bf 1008}, 107 (2010).
[arXiv:1003.4244 [hep-th]].
}

\lref\deBult{
  F.~J.~van~de~Bult,
  ``An elliptic hypergeometric integral with $W(F_4)$ symmetry,''
The Ramanujan Journal, Volume 25, Issue 1 (2011)
[arXiv:0909.4793[math.CA]].
}

\lref\GaddeKB{
  A.~Gadde, E.~Pomoni, L.~Rastelli and S.~S.~Razamat,
  ``S-duality and 2d Topological QFT,''
JHEP {\bf 1003}, 032 (2010).
[arXiv:0910.2225 [hep-th]].
}

\lref\ArgyresCN{
  P.~C.~Argyres and N.~Seiberg,
  ``S-duality in N=2 supersymmetric gauge theories,''
JHEP {\bf 0712}, 088 (2007).
[arXiv:0711.0054 [hep-th]].
}

\lref\SpirWarnaar{
  V.~P.~Spiridonov and S.~O.~Warnaar,
  ``Inversions of integral operators and elliptic beta integrals on root systems,''
Adv. Math. 207 (2006), 91-132
[arXiv:math/0411044].
}

\lref\GaiottoHG{
  D.~Gaiotto, G.~W.~Moore and A.~Neitzke,
  ``Wall-crossing, Hitchin Systems, and the WKB Approximation,''
[arXiv:0907.3987 [hep-th]].
}

\lref\RuijsenaarsVQ{
  S.~N.~M.~Ruijsenaars and H.~Schneider,
  ``A New Class Of Integrable Systems And Its Relation To Solitons,''
Annals Phys.\  {\bf 170}, 370 (1986).
}

\lref\RuijsenaarsPP{
  S.~N.~M.~Ruijsenaars,
  ``Complete Integrability Of Relativistic Calogero-moser Systems And Elliptic Function Identities,''
Commun.\ Math.\ Phys.\  {\bf 110}, 191 (1987).
}

\lref\HallnasNB{
  M.~Hallnas and S.~Ruijsenaars,
  ``Kernel functions and Baecklund transformations for relativistic Calogero-Moser and Toda systems,''
J.\ Math.\ Phys.\  {\bf 53}, 123512 (2012).
}

\lref\kernelA{
S.~Ruijsenaars,
  ``Elliptic integrable systems of Calogero-Moser type: Some new results on joint eigenfunctions'', in Proceedings of the 2004 Kyoto Workshop on "Elliptic integrable systems", (M. Noumi, K. Takasaki, Eds.), Rokko Lectures in Math., no. 18, Dept. of Math., Kobe Univ.
}

\lref\ellRSreview{
Y.~Komori and S.~Ruijsenaars,
  ``Elliptic integrable systems of Calogero-Moser type: A survey'', in Proceedings of the 2004 Kyoto Workshop on "Elliptic integrable systems", (M. Noumi, K. Takasaki, Eds.), Rokko Lectures in Math., no. 18, Dept. of Math., Kobe Univ.
}

\lref\langmann{
E.~Langmann,
  ``Explicit solution of the (quantum) elliptic Calogero-Sutherland model'', [arXiv:math-ph/0401029].
}

\lref\langmannB{
E.~Langmann,
  ``An explicit solution of the (quantum) elliptic Calogero-Sutherland model'', [arXiv:math-ph/0407050].
}

\lref\langmannC{
E.~Langmann,
  ``Conformal field theory and the solution of the (quantum) elliptic Calogero-Sutherland system'', [arXiv:math-ph/0411081].
}

\lref\TachikawaWI{
  Y.~Tachikawa,
  ``4d partition function on $S^1 \times S^3$ and 2d Yang-Mills with nonzero area,''
PTEP {\bf 2013}, 013B01 (2013).
[arXiv:1207.3497 [hep-th]].
}

\lref\MinahanFG{
  J.~A.~Minahan and D.~Nemeschansky,
  ``An N=2 superconformal fixed point with E(6) global symmetry,''
Nucl.\ Phys.\ B {\bf 482}, 142 (1996).
[hep-th/9608047].
}

\lref\AldayKDA{
  L.~F.~Alday, M.~Bullimore, M.~Fluder and L.~Hollands,
  ``Surface defects, the superconformal index and q-deformed Yang-Mills,''
[arXiv:1303.4460 [hep-th]].
}

\lref\FukudaJR{
  Y.~Fukuda, T.~Kawano and N.~Matsumiya,
  ``5D SYM and 2D q-Deformed YM,''
Nucl.\ Phys.\ B {\bf 869}, 493 (2013).
[arXiv:1210.2855 [hep-th]].
}

\lref\XieHS{
  D.~Xie,
  ``General Argyres-Douglas Theory,''
JHEP {\bf 1301}, 100 (2013).
[arXiv:1204.2270 [hep-th]].
}

\lref\prodMac{
M.~Rahman, A.~Verma,
  ``Product and Addition Formulas for the Continuous q-Ultraspherical Polynomials,''
SIAM J. Math. Anal., 17(6).
}

\lref\GorskyDQ{
 A.~Gorsky and N.~Nekrasov,
 ``Relativistic Calogero-Moser model as gauged WZW theory,''
Nucl.\ Phys.\ B {\bf 436}, 582 (1995).
[hep-th/9401017].
}

\lref\GorskyDJ{
  A.~Gorsky and N.~Nekrasov,
  ``Elliptic Calogero-Moser system from two-dimensional current algebra,''
[hep-th/9401021].
}

\lref\NekrasovCZ{
  N.~Nekrasov,
  ``Five dimensional gauge theories and relativistic integrable systems,''
Nucl.\ Phys.\ B {\bf 531}, 323 (1998).
[hep-th/9609219].
}

\lref\NekrasovRC{
  N.~A.~Nekrasov and S.~L.~Shatashvili,
  ``Quantization of Integrable Systems and Four Dimensional Gauge Theories,''
[arXiv:0908.4052 [hep-th]].
}

\lref\GorskyPX{
  A.~Gorsky and A.~Mironov,
  ``Integrable many body systems and gauge theories,''
In *Aratyn, H. (ed.) et al.: Integrable hierarchies and modern physical theories* 33-176.
[hep-th/0011197].
}

\lref\GaiottoBWA{
  D.~Gaiotto and P.~Koroteev,
``On Three Dimensional Quiver Gauge Theories and Integrability,''
JHEP {\bf 1305}, 126 (2013).
[arXiv:1304.0779 [hep-th]].
}

\Title{\vbox{\baselineskip12pt
}}
{\vbox{
\centerline{On the ${\cal N}=2$ superconformal index}
\vskip7pt 
\centerline{and eigenfunctions of the elliptic RS model}
}
}

\centerline{Shlomo S. Razamat}
\bigskip
\centerline{{\it School of Natural Sciences, Institute for Advanced Study, Princeton, NJ 08540, USA}}
\vskip.1in \vskip.2in \centerline{\bf Abstract}

\vskip.2in

\noindent
We define an infinite sequence of superconformal indices, ${\cal I}_n$, generalizing the Schur index for
${\cal N}=2$ theories. For theories of class ${\cal S}$ we then suggest a recursive technique
to completely determine ${\cal I}_n$.  The information encoded in the sequence 
of indices is equivalent to the ${\cal N}=2$ superconformal index depending on a maximal set 
of fugacities. Mathematically, the  procedure suggested in this note 
provides a perturbative algorithm for computing a set of eigenfunctions of the
elliptic Ruijsenaars-Schneider model.

\vfill

\Date{August 2013}


\newsec{Introduction}

It often happens that some of the properties of a given physical system can be 
constrained by exploiting the symmetries, and other robust properties, of that system.
 In  more rare circumstances some 
of the observables are completely determined by constraints imposed by robust  considerations.
Such observables are  especially useful if, say due to strong coupling, direct 
computations in the theory describing the physical system are not feasible.

An example of a quantity fixed by symmetries and robust properties is the superconformal index 
of the ${\cal N}=2$ quantum field theories of class ${\cal S}$~\refs{\GaiottoWE,\GaiottoHG}. The superconformal index is a twisted supersymmetric 
 partition function on $\S^3\times \S^1$ and thus is severely constrained by supersymmetry.
Moreover, due to an intricate web of  relations between different theories of class ${\cal S}$,
 S-dualities and RG flows, under certain mathematical assumptions to be discussed below, the
physical task of determining the superconformal index 
can be reduced to  a well defined mathematical problem of diagonalizing a  set 
of commuting difference operators~\GaiottoXA. The purpose of this brief note is to argue 
that this mathematical problem itself can be solved by exploiting further the S-dualities
relating the theories of class ${\cal S}$.

The difference operators one has to diagonalize in order to determine the 
superconformal index are the elliptic Ruijsenaars-Schneider (RS) Hamiltonians~\refs{\RuijsenaarsVQ,
\RuijsenaarsPP,\ellRSreview}. We will see that S-duality properties relating different theories of class
${\cal S}$ of type $A_{N-1}$ can be translated to a system of coupled integral equations
for the eigenfunctions of elliptic RS Hamiltonians (and additional un-known functions).
A solution of this system of equations is known in closed form in particular limits of  the parameters.
We will then construct the eigenfunctions for the general case by 
perturbing around one such limit, the so called Schur index~\refs{\GaddeIK,\GaddeUV}.

The Schur index has several interesting properties: {\it e.g.}, it is related to qYM in 
 2d~\refs{\GaddeIK,\TachikawaWI,\FukudaJR,\AldayKDA},
it has interesting modular properties and is given in terms of q-polyGamma functions~\RazamatUV.
This version of the index depends only on one fugacity, which we denote by $q$,
 coupled to charges of the superconformal
symmetry group. This is to be contrasted with the most general 4d ${\cal N}=2$ index one can define depending on three superconformal fugacities~\KinneyEJ,
 usually denoted $(p,q,t)$: the Schur index is obtained by setting $t=q$. We will define
the general ${\cal N}=2$ index as a power series in $T=(t/q)^{\frac12}-1$. The coefficients
of this series are indices by themselves, which we will denote by ${\cal I}_n$.  We then solve for 
the eigenfunctions of elliptic RS as Taylor series in $T$.  Knowing these eigenfunctions up to order $\widetilde n$ in $T$ is sufficient to determine the indices ${\cal I}_{n\leq \widetilde n}$.

\

Let us mention here that the interplay between quantum mechanical integrable models and
gauge theories in higher dimensions has a rich history.  For example, an explicit  relation
between the trigonometric RS model and gauged $G/G$ WZW model
was discussed in \refs{\GorskyDQ}. In~\GorskyDJ\ a generalization of this relation
to elliptic non-relativistic models appeared, while  the elliptic RS model was found to be connected to 5d gauge theories on a circle~\NekrasovCZ.
For a nice review of these, and other, relations and more references one can consult~\refs{\GorskyPX}.
More recently, the quantum elliptic RS model made its appearance in~\NekrasovRC\
while considering  ${\cal N}=2$ theories in the $\Omega$-background.\foot{
It would be interesting to work out in detail the relations between these  
appearances of RS models in ${\cal N}=2$ gauge theories
 and the index: see {\it e.g.},~\GaiottoBWA\  
for a suggestion.}

\

This note is organized as follows. We start in section 1 with a definition of a sequence of
${\cal N}=2$ indices, ${\cal I}_n$, generalizing the Schur index. In section 2 we discuss $A_1$ theories 
of class ${\cal S}$ and spell out a simple algorithm to determine a set of eigenfunctions 
of the elliptic RS model, which are needed to compute ${\cal I}_n$. In section 3, as a concrete example
of analysis for $A_{N-1}$ theories of class ${\cal S}$ with $N>2$, we extend 
the algorithm to the $A_2$ case. Finally, in section 4 we briefly comment on our results. 

\

Before proceeding to the main part of the note let us set several  mathematical notations. The elliptic 
Gamma function is defined as,

\eqn\ellGamma{
\Gamma(x;p,\,q)=\prod_{i,\,j\geq 0} \frac{1-\frac{p\,q}{x}\;p^i q^j}{1-x\;p^i q^j}\,.
} The q-Pochhammer symbol and the theta-function are given by,

\eqn\pochAndtheta{
(x;\,q) = \prod_{i\geq0} (1-x\;q^i),\quad\qquad
\theta(x;\,q)=(x;\,q)\;(\frac{q}{x};\,q)\,,
}  and we will find it useful to define the following

\eqn\kappadef{
\kappa_n=\frac{\left((q;q)(p;p)\Gamma(\frac{p\,q}{t};p,q)\right)^n}{(n+1)!}\,.
} Finally, we use the usual short-hand notation,
\eqn\short{
f(z^{\pm1}y^{\pm1}\cdots) = f(z y\cdots)\;f(z^{-1} y\cdots)\;f(z y^{-1}\cdots)\;f(z^{-1} y^{-1}\cdots)\,.
}

\

\newsec{A sequence of indices}

The ${\cal N}=2$ superconformal index is given by the following trace formula, 

\eqn\index{
{\cal I}(p,q,t,\{a_\ell\})=\Tr(-1)^F p^{j_2+j_1-r}\;q^{j_2-j_1-r}\;t^{r+R}\;e^{-\beta\,\delta}\,\prod_{\ell} a_\ell^{f_\ell}\,.
} The trace is over the states of a $4d$ ${\cal N}=2$ superconformal theory in  radial quantization.
Here $F$ is the fermion number, $(j_1, j_2)$ the Cartans of the rotation group $SU(2)_1\times
SU(2)_2\sim SO(4)$, $R$ is the Cartan of $SU(2)_R$ R-symmetry, 
$r$ the generator of $U(1)_r$  R-symmetry, and $f_\ell$ are Cartans of the flavor symmetry (if there is any).
The charge $\delta$ is an anti-commutator of two supercharges,\foot{
We use the notations of~\GaddeUV.}

\eqn\deltadef{
\delta=2\left\{\widetilde {\cal Q}_{1,\dot{-}},\,\left(\widetilde {\cal Q}_{1,\dot{-}}\right)^\dagger\right\}=
E-2j_2-2R+r\,.
} All the charges in \index\ commute with $\widetilde {\cal Q}_{1,\dot{-}}$ and thus the index is 
independnet of $\delta$: we will omit it from the discussion in what follows. The supercharge
$\widetilde {\cal Q}_{1,\dot{-}}$ has the following charges $(R,r,j_1,j_2)=(\frac12,-\frac12,0,-\frac12)$.

\

\noindent We  introduce a new parameter $T$,

\eqn\rdef{
1+T=(t/q)^{1/2}\,,
} and define a sequence of indices, ${\cal I}_n$, by the following 
\eqn\expT{
{\cal I}=\sum_{n=0}^\infty T^n\;{\cal I}_n\,.
} 
In all cases where the index can be directly computed such an expansion in $T$ is convergent and 
we  assume this to be true for all theories we will discuss.
The indices ${\cal I}_n$ have a simple physical meaning

\eqn\indexlead{
{\cal I}_n=\Tr(-1)^F P_n(r+R)\,p^{j_2+j_1-r}\;q^{j_2-j_1+R}\prod_{\ell} a_\ell^{f_\ell}\,,
} with $P_0(z)=1$ and $P_n(z)=\frac{1}{n!}\prod_{\ell=0}^{n-1}(2z-\ell)$.
The index ${\cal I}_0$ is
the Schur index~\refs{\GaddeIK,\GaddeUV},

\eqn\schurind{
{\cal I}_0=\Tr(-1)^F \,p^{j_2+j_1-r}\;q^{j_2-j_1+R}\prod_{\ell} a_\ell^{f_\ell}=
\Tr(-1)^F \,q^{j_2-j_1+R}\prod_{\ell} a_\ell^{f_\ell}\,.
} Here the dependence on $p$ completely drops out since the charge it couples to
is (up to admixture of $\delta$) an anticommutator of a supercharge, ${\cal Q}_{1,+}$ (with charges $(R,r,j_1,j_2)=(\frac12,\frac12,\frac12,0)$), which commutes with all the other charges appearing in the index.
The indices ${\cal I}_{n>0}$, however, do depend on $p$ since ${\cal Q}_{1,+}$ has non vanishing $r+R$ charge, {\it e.g.}

\eqn\indexleadOne{
{\cal I}_1=\Tr(-1)^F 2(r+R)\,p^{j_2+j_1-r}\;q^{j_2-j_1+R}\prod_{\ell} a_\ell^{f_\ell}\,.
}
The coefficients of the term $p^nq^m\prod_{\ell} a_\ell^{f_\ell}$ in expansion of ${\cal I}_1$
 give a graded sum over R-charges of states
with $j_2+j_1-r=n$, $j_2-j_1+R=m$, $\delta=0$, and the flavor charges $f_\ell$.
Note that the indices ${\cal I}_n$ do not have the symmmetry exchanging $p$ and $q$.

Using the algorithm of this note one can compute the indices ${\cal I}_n$ for theories of class ${\cal S}$. We will
give explicit ingredients for the computation of ${\cal I}_{1}$ for $A_1$ quivers
and $A_2$ quivers.

\

\newsec{The intergral equation: $A_1$ case}

Theories of class ${\cal S}$ of type $A_{N-1}$ are obtained by compactifying $M5$-branes on a punctured Riemann surface~\refs{\GaiottoWE,\GaiottoHG}. These are $4d$ ${\cal N}=2$ theories  labeled by 
the choice of the punctured Riemann surface. The modular parameters of the Riemann surface 
correspond to exactly marginal couplings of the $4d$ theory. The superconformal index is independent
of such continuous parameters of the theory and thus can be thought of as a topological invariant
associated to the underlying Riemann surface~\GaddeKB. The types of punctures are classified by partitions of $N$ and encode information about the flavor  symmetry of the $4d$ theory. 

In this section we discuss the simplest case of theories of class ${\cal S}$,  the $A_1$ generalized quivers.
For $A_1$ 
case there is only one type of non-trivial  puncture corresponding to an $SU(2)$ flavor symmetry.
The index of a class ${\cal S}$ theory of $A_1$ type corresponding to genus 
$g$ Riemann surface with $s$  punctures\foot{We only discuss regular punctures in this note~\XieHS.} can be constructed
as follows.  The index of a theory corresponding to three-punctured sphere (trinion) is
\eqn\trinionAone{
{\cal I}^{(0,3)}(a,b,c)=\Gamma(t^{1/2}a^{\pm1}b^{\pm1}\,c^{\pm1};p,q)\,.
} Here the parameters $a$, $b$, and $c$ are fugacities for the Cartans of the three $SU(2)$ 
flavor symmetries associated to the three punctures. 
The corresponding $4d$ theory is a free hyper-multiplet in the bi-fundamental  representation
 of $SU(2)\times SU(2)$.
 Given two general Riemann
surfaces and the corresponding indices one constructs the index of the theory associated 
to the Riemann surface obtained by gluing the two surfaces along a puncture,

\eqn\gluingriemann{
\eqalign{
&{\cal I}^{(g+g',s+s'-2)}(\{a_i\}_{i=1}^{s+s'-1})=\cr
&\qquad\kappa_1\oint\frac{dz}{2\pi i z}
\frac{\Gamma(\frac{pq}{t}z^{\pm2};p,q)}
{\Gamma(z^{\pm2};p,q)}\;{\cal I}^{(g,s)}(z,\{a_i\}_{i=1}^{s-1})\;
{\cal I}^{(g',s')}(z^{-1},\{a_i\}_{i=s}^{s+s'-2})\,.
}
}  Gluing two Riemann surfaces corresponds to gauging  the diagonal combination of the $SU(2)$ 
flavor symmetries associated to the glued punctures.
The ratio of two elliptic Gamma functions and $\kappa_1$ in~\gluingriemann\ come from the index of the ${\cal N}=2$ vector multiplet (and the Haar measure) one introduces when a symmetry is gauged.
The index for any Riemann surface is constructed recursively by decomposing the surface into
three-punctured spheres (pairs-of-pants) and gluing them together. 

\

Our perturbative procedure to determine the indices ${\cal I}_n$ is based on two main assumptions: 
one physical and one mathematical. 
The physics we will assume to be correct is that the theories of class ${\cal S}$ enjoy the action
of S-duality as discussed in~\GaiottoWE. In particular this means that the indices computed
in any duality frame are equal~\GaddeKB:
this  statement is a mathematically proven fact at the level of the index
for the $A_1$ theories of class ${\cal S}$ (with regular punctures)~\deBult.
Saying it differently, S-duality is the assumption that the recursive procedure defined
above to construct indices of theories associated to  general Riemann surfaces is well defined, {\it i.e.},
independent of the pairs-of-pants decomposition.

\

The main mathematical assumption we will make is that the index
of a theory of class ${\cal S}$ can be written in the following form,\foot{
Here we assume that for $g=0$ the number of punctures is bigger than two, and for $g=1$ there is
at least one puncture.
}

\eqn\mathass{
{\cal I}^{(g,s)}(p,q,t;\{a_\ell\})=
\sum_{\lambda=0}^\infty \left[{\cal C}_\lambda\right]^{2g-2+s}
\prod_{i=1}^s\psi_\lambda(a_i;p,q,t)\,,
} where the functions $\psi_\lambda(a;p,q,t)$ form an orthonormal set of functions under the vector multiplet measure,\foot{We will often omit the parameters $(p,q,t)$ for the sake of brevity.}

\eqn\measure{
\langle\langle\psi_\lambda,\psi_\mu\rangle\rangle=
\kappa_1\oint\frac{dz}{2\pi i z}
\frac{\Gamma(\frac{pq}{t}z^{\pm2};p,q)}
{\Gamma(z^{\pm2};p,q)}\;
\psi_\lambda(z)\;
\psi_\mu(z)=\delta_{\lambda,\mu}\,.
}
One can use the physical assumption above to motivate the expression \mathass\ ~\GaddeUV.
 Take the  index 
of the $A_1$ trinion~\trinionAone\ and expand it in some convenient complete set of functions,
$f_\mu(z)$, orthonormal under the 
vector multiplet measure \measure. Such an expansion defines a set of symmetric structure constants,
$C_{\mu\nu\lambda}$. These structure constants have to form an associative algebra
following from the S-duality properties of the underlying theories.  Then pending certain natural assumptions and convergence issues one can change a basis of the orthogonal functions such that
the structure constants will be ``diagonalized'' to Frobenius form \mathass. 
In such a ``diagonal'' form the S-duality properties of the indices of $A_1$ theories are manifest.
At least when one specializes the parameters $(p,q,t)$ to  certain values there is a lot of 
 evidence that the mathematical assumption \mathass\ is correct. For example,
when one specializes the fugacities to $p=0$ (or $q=0$) the functions $\psi_\lambda(a)$ 
can be explicitly found to be related to Macdoanld polynomials~\GaddeUV, and taking $t=q$ these are related to Schur polynomials
as we will discuss in the next sub-section.  
If the functions $\psi_\lambda(a)$ are known, utilizing \mathass\ to compute the indices
of general $A_1$ quivers is much simpler than applying the recursive procedure using
\gluingriemann.  We thus turn to deriving our basic integral equation determining the eigenfunctions
$\psi_\lambda(a)$.

\

It will be convenient for us to absorb a factor of ${\cal C}_\lambda$ into the definition of the functions,

\eqn\hatpsi{
\hat \psi_\lambda(z;p,q,t) = 
{\cal C}_\lambda\;{\psi_\lambda(z;p,q,t)}
\,.
}
The basic example of an index, the index of the theory corresponding to the three-punctured sphere~\trinionAone, can be then written as
\eqn\trinioind{
{\cal I}^{(0,3)}=\Gamma(t^{1/2}a^{\pm1}b^{\pm1}\,c^{\pm1};p,q)=
\sum_{\lambda=0}^\infty
{\cal C}_\lambda^{-2}\;
\hat \psi_\lambda(a;p,q,t)
\hat \psi_\lambda(b;p,q,t)
\hat \psi_\lambda(c;p,q,t)
\,.
} Using the orthogonality property \measure\ we immediately arrive at our basic integral equation
for $\hat \psi_\lambda(z;p,q,t)$,\foot{
This is an elliptic generalization of the product formula for $A_1$ Macdonald polynomials~\prodMac.
}

\eqn\eigenres{
\eqalign{
&\hat \psi_\lambda(y;p,q,t)\;\hat \psi_\lambda(a;p,q,t)=\cr
&\qquad\kappa_1\oint\frac{dz}{2\pi i z}
\frac{\Gamma(\frac{pq}{t}z^{\pm2};p,q)}
{\Gamma(z^{\pm2};p,q)}\; \hat \psi_\lambda(z;p,q,t)\;
\Gamma(t^{1/2}z^{\pm1}y^{\pm1}\,a^{\pm1};p,q)\,.
}
} 
In the next sub-section we will argue that given the particular solution of this equation for $t=q$,
it uniquely fixes $\hat \psi_\lambda(z;p,q,t)$ for $t\neq q$.

\

Let us now comment on the relation of the above to the elliptic Ruijsenaars-Schneider model.
It has been argued in~\GaiottoXA\ that there is a set of difference operators, ${\frak S}^{(p,q,t)}_{(m,n)}$,
for which the indices ${\cal I}^{(g,s)}$ are {\it{kernel functions}} (see {\it e.g.}~\refs{\langmann,\kernelA,\HallnasNB,\noumi} for discussion of kernel functions),\foot{
Kernel functions were instrumental in the perturbative solution of the non-relativistic elliptic Calogero-Moser
model~\refs{\langmann,\langmannC}.
}

\eqn\kernelfunc{
{\frak S}^{(p,q,t)}_{(m,n)}(a_\alpha)\cdot {\cal I}^{(g,s)}(\{a_i\})=
{\frak S}^{(p,q,t)}_{(m,n)}(a_\beta)\cdot {\cal I}^{(g,s)}(\{a_i\})\,, \qquad \forall\; \alpha\,, \;\beta \in \{1,2,\cdots
s\}\,.
}
These operators can be found in~\GaiottoXA\ and here we mention explicitly only the most basic one,

\eqn\diffOP{
{\frak S}^{(p,q,t)}_{(1,0)}(\{z_\ell\})\cdot f(\{z_\ell\})=
\sum_{i=1}^N\prod_{j\neq i}\frac{\theta(\frac{t}{p} z_i/z_j;q)}{\theta( z_j/z_i;q)}\;
f(\{z_\ell\to p^{\frac{1}N-\delta_{\ell\,i}} z_\ell\})\,.
} We wrote this operator for $A_{N-1}$ case, so for $A_1$ one just has to set $N=2$ in the above.
The difference operators ${\frak S}^{(p,q,t)}_{(0,1)}$ is obtained by exchanging $p$ and $q$, ${\frak S}^{(p,q,t)}_{(0,1)}={\frak S}^{(q,p,t)}_{(1,0)}$. The operator \diffOP\ is,
up to conjugation with a simple function~\GaiottoXA, the basic Hamiltonian of the elliptic Ruijsenaars-Schneider  model.
The operators \diffOP\ are self-adjoint under the natural measure in our problem, eq.~\measure.

The property \kernelfunc\ together with the assumption \mathass\ and
the orthogonality \measure\ imply that the functions
$\hat \psi_\lambda(z;p,q,t)$ are eigenfunctions of \diffOP\ and its generalizations
appearing in~\GaiottoXA. For example take ${\cal I}^{(1,3)}$ and act on it with \diffOP,
\eqn\condRS{\eqalign{
&\sum_{\lambda=0}^\infty\left[{\frak S}^{(p,q,t)}_{(1,0)}(a_1)\hat \psi_\lambda(a_1;p,q,t)\right]
\,\prod_{i=2,3}\hat \psi_\lambda(a_i;p,q,t)=\cr
&\qquad\qquad\qquad\qquad
\sum_{\lambda=0}^\infty\left[{\frak S}^{(p,q,t)}_{(1,0)}(a_2)\hat \psi_\lambda(a_2;p,q,t)\right]
\,\prod_{i=1,3}\hat \psi_\lambda(a_i;p,q,t)
\,,\cr&\qquad\qquad\qquad\qquad\qquad\qquad\downarrow\qquad\qquad\cr
&\frac{{\frak S}^{(p,q,t)}_{(1,0)}(a_1)\hat \psi_\lambda(a_1;p,q,t)}{\hat \psi_\lambda(a_1;p,q,t)}=
\frac{{\frak S}^{(p,q,t)}_{(1,0)}(a_2)\hat \psi_\lambda(a_2;p,q,t)}{\hat \psi_\lambda(a_2;p,q,t)}\,.
}
} Going from first to second equality we applied orthogonality of the eigenfunctions
$\hat \psi_\lambda(a_3;p,q,t)$.
Here we have also assumed that the order of performing the infinite sum and acting with the difference
operator is immaterial.
In degeneration limits, say $p=0$ or $t=q$, the functions $\hat \psi_\lambda(z;p,q,t)$
can be explicitly shown to be eigenfunctions of \diffOP.  
 In what follows we will give some evidence that
indeed this is the case beyond $t=q$, and in particular that  our mathematical 
and physical assumptions are consistent.

\

\subsec{Recursive solution}

Let us now turn to solving the integral equations~\eigenres\ perturbatively starting with a known solution
for specialized parameters, $t=q$. 
We set the small parameter $T$ to be as in~\rdef,
$$
1+T=(t/q)^{1/2}\,.
$$
 We define
\eqn\kernel{
\Delta(z,y,a;p,q,t)=\kappa_1\, \frac{\Gamma(\frac{pq}{t}z^{\pm2};p,q)}
{\Gamma(z^{\pm2};p,q)}\;
\Gamma(t^{1/2}z^{\pm1}y^{\pm1}\,a^{\pm1};p,q)=
\sum_{n=0}^\infty T^n\; \Delta^{(n)}(z,y,a;p,q)\,,
}
and
\eqn\pertfunc{
\hat \psi_\lambda(z;p,q,t)=
\sum_{n=0}^\infty T^n\;\hat \psi^{(n)}_\lambda(z;p,q)=
\sum_{n=0}^\infty T^n\;\sum_{\mu=0}^\infty C_{\mu;\;\lambda}^{(n)}(p,q)\;\hat \psi^{(0)}_\mu(z;q)\,.
} Here $\hat\psi^{(0)}_\mu(z;q)$ is the solution to ~\eigenres\ in the Schur case, $t=q$, which is known~\refs{\GaddeIK,\GaddeUV},

\eqn\schursol{
\hat\psi^{(0)}_\mu(z;q)=
\frac{1}{1-q}
\left(
\prod_{\ell=0}^\infty\frac{1}{1-q^{1+\ell} \, z^{\pm2}}
\right)\;
\frac{\chi_\mu(z)}{\chi_\mu(q^{\frac12})}\,,\qquad\qquad
\chi_\mu(z)=\frac{z^{\mu+1}-z^{-\mu-1}}{z-z^{-1}}\,.
} 
These functions are eigenfunctions of difference operators~\diffOP\ when one takes $t=q$,

\eqn\eigenszero{
{\frak S}^{(p,q,q)}_{(1,0)}\cdot \hat\psi^{(0)}_\mu=(p^{-\frac{\mu}2}+p^{\frac{\mu}2+1})\;\hat\psi^{(0)}_\mu\,,\qquad\qquad
{\frak S}^{(p,q,q)}_{(0,1)}\cdot \hat\psi^{(0)}_\mu=(q^{-\frac{\mu}2}+q^{\frac{\mu}2+1})\;\hat\psi^{(0)}_\mu\,.
}
The coefficients $C_{\mu;\;\lambda}^{(n)}(q,p)$ are given by
\eqn\coefCs{
C_{\mu;\;\lambda}^{(n)}(p,q)=\frac{\langle \hat \psi^{(n)}_\lambda(z;p,q),\,\hat \psi^{(0)}_\mu(z;q) \rangle}{\langle \hat \psi^{(0)}_\mu(z;q),\, \hat \psi^{(0)}_\mu(z;q)\rangle}\,,
} with the inner product defined using the specialization of the vector multiplet 
measure~\measure\ to the Schur case,
\eqn\schurmeas{
<f,g>=\oint \frac{dz}{2\pi i z}\Delta(z;q)f(z)g(z)\,,\qquad\qquad \Delta(z;q)=
\frac{(q;q)^2}2 \;\theta(q\,z^2;q)\;\theta(q\,z^{-2};q)\,.
}

Using these definitions the integral equation \eigenres\ becomes an equation for $C^{(n)}_{\nu;\;\lambda}$. Taking $n>0$ one obtains
\eqn\solsA{
\eqalign{
&C^{(n)}_{\nu;\;\lambda}\delta_\mu^\lambda+
C^{(n)}_{\mu;\;\lambda}\delta_\nu^\lambda+
\sum_{k+l=n,\, k,l<n}C^{(k)}_{\nu;\;\lambda}C^{(l)}_{\mu;\;\lambda}=
C^{(n)}_{\mu;\;\lambda}\,\delta_{\mu\nu}+
\sum_{\rho=0}^\infty\sum_{k+l=n,\,l<n}{\Gamma^{(k)}}^{\rho}_{\mu\nu}\;
C^{(l)}_{\rho;\;\lambda}\,,
}
} where,
\eqn\Gmmadef{
\eqalign{
&{\Gamma^{(k)}}^{\rho}_{\mu\nu}=\cr
&\oint \frac{dz}{2\pi i z}
\oint \frac{dy}{2\pi i y}
\oint \frac{da}{2\pi i a}
\frac{\Delta^{(k)}(z,y,a;p,q)\;
\Delta(y;q)\;\Delta(a;q)\;
\hat \psi^{(0)}_\rho(z;q)\hat \psi^{(0)}_\mu(y;q)\hat \psi^{(0)}_\nu(a;q)}
{\langle \hat \psi^{(0)}_\mu,\, \hat \psi^{(0)}_\mu\rangle
\langle \hat \psi^{(0)}_\nu,\, \hat \psi^{(0)}_\nu\rangle}\,,
}
} with $\Delta(z;q)$, defined in \schurmeas, being the measure under which the  functions~\schursol\ are orthogonal.

\

\noindent One can now recursively solve these equations. Taking $n=1$ we obtain immediately
\eqn\nonesol{
\eqalign{
&C^{(1)}_{\mu;\;\lambda}={\Gamma^{(1)}}^{\lambda}_{\mu\lambda}=
-{\Gamma^{(1)}}^{\lambda}_{\mu\mu}\quad (\lambda\neq \mu)\,,\qquad\qquad
{\Gamma^{(1)}}^{\rho}_{\mu\nu}=0\quad (\rho\neq\mu,\;\rho\neq\nu,\;\mu\neq \nu)\,,\cr
&C^{(1)}_{\mu;\;\mu}={\Gamma^{(1)}}^{\mu}_{\mu\mu}\,.
}
} Note that the set of equations is overdetrmined and there are thus consistency conditions
to be satisfied: for the procedure to have a solution the functions ${\Gamma^{(1)}}^{\rho}_{\mu\nu}$
are not arbitrary. Similarly, at the second level, $n=2$, one obtains
\eqn\ntwosol{
\eqalign{
&C^{(2)}_{\mu;\;\lambda}=C^{(1)}_{\lambda;\;\mu}\,C^{(1)}_{\mu;\;\lambda}+{\Gamma^{(2)}}^{\lambda}_{\mu\lambda}=
\left(C^{(1)}_{\mu;\;\lambda}\right)^2+
\sum_{\rho=0}^\infty (-1)^{\delta_{\mu\rho}}C^{(1)}_{\mu;\;\rho}\;C^{(1)}_{\rho;\;\lambda}
-{\Gamma^{(2)}}^{\lambda}_{\mu\mu}\quad (\lambda\neq \mu)\,,\cr
&{\Gamma^{(2)}}^{\rho}_{\mu\nu}=
C^{(1)}_{\mu;\;\rho}\,C^{(1)}_{\nu;\;\rho}-
C^{(1)}_{\mu;\;\rho}\,C^{(1)}_{\nu;\;\mu}-
C^{(1)}_{\nu;\;\rho}\,C^{(1)}_{\mu;\;\nu}\,
\quad (\rho\neq\mu,\;\rho\neq\nu,\;\mu\neq \nu)\,,\cr
&C^{(2)}_{\mu;\;\mu}=
{\Gamma^{(2)}}^{\mu}_{\mu\mu}-\sum_{\rho=0,\rho\neq\mu}^\infty C^{(1)}_{\mu;\;\rho}\,
C^{(1)}_{\rho;\;\mu}\,.
}
} And so on.
As an exercise one can set $p=0$ and verify that the recursive procedure 
generates the known solutions~\GaddeUV,

\eqn\macsol{
\hat \psi_\mu(z;p=0,q,t)=
\frac{(t^2;q)}{(t;q)}
\left(
\prod_{\ell=0}^\infty\frac{1}{1-t\,q^{\ell} \, z^{\pm2}}
\right)\;
\frac{P_\mu(z;q,\,t)}{P_\mu(t^{\frac12};q,\,t)}\,,
} where $P_\mu(z;q,\,t)$ are the Macdonald polynomials.

Our goal was to  generate the solution for the elliptic case, and here are several eigenfunctions
up to leading correction in $T$, and expanded to several orders in $p$ and $q$,\foot{We performed explicit computations using {\it{Mathematica}}.}

\eqn\eigebfuncslist{
\eqalign{
&\hat\psi_0(z)=
1 + q + 2 q^2 + T\,(2 q - 2 p^2   + 4 p q  + 2 p^2 q  + 6 q^2  + 
 8 p q^2  + 8 p^2 q^2 )+\cr
&\qquad (z^2+z^{-2})(q+2 q^2+T\,(2 q +2 p q +2 p^2 q +6 q^2 +6 p q^2 +6 p^2 q^2))+\cr
&\qquad (z^4+z^{-4})(q^2+T\,(4 q^2 +2 p q^2 +4 p^2 q^2 ))+\cdots\,,\cr
&\hat\psi_1(z)=(z+z^{-1})(q^{1/2}+q^{3/2}+T(q^{1/2} +q^{3/2} (3 +4 p +4 p^2 )))+\cr
&\qquad\qquad (z^3+z^{-3})(q^{3/2} (1 + T\,(3  + 2 p  + 2 p^2 )))+\cdots\,,\cr
&\hat\psi_2(z)=q + 2 q^2 + T\,(2 q  + 2 p q  + 6 q^2  + 4 p q^2  + 4 p^2 q^2 )+\cr
&\qquad\qquad (z^2+z^{-2})(q+q^2+T\,(2 q +4 q^2 +4 p q^2 +4 p^2 q^2 ))+\cr
&\qquad\qquad  (z^4+z^{-4})(q^2+T\,(4 q^2 +2 p q^2 +2 p^2 q^2 ))+\cdots\,,\cr
&\hat\psi_3(z)=(z+z^{-1})\,q^{3/2} (1 + T(3  + 2 p  + 2 p^2 ))+
(z^3+z^{-3})q^{3/2} (1+3 T)+\cdots\,.
}
}
One can check that these eigenfunctions are orthogonal under the vector multiplet measure~\measure\ and compute the eigenvalues.
For example the eigenvalues of the $\lambda=0,1$ eigenfunctions with respect to the operator
${\frak S}_{(1,0)}^{(p,q,t)}$ are

\eqn\eigenZero{
\eqalign{
&{\cal E}_{(1,0)}^{\lambda=0}=
1+p+2\,T\,(q+q^2-p-pq-pq^2+\dots)\,,\cr
&{\cal E}_{(1,0)}^{\lambda=1}=p^{-\frac12}+p^{\frac32}+2\,T\,(p^{-\frac12}q^2-p^{\frac{3}{2}})\,+\cdots\,.
}
} From the general structure of the difference operator \diffOP\ it is possible  to deduce that the (orthonormal) eigenfunctions have to satisfy~\GaiottoXA

\eqn\reltpqt{
\psi_\lambda(z;p,q,t)=\Gamma(t;p,q)\,\Gamma(tz^2;p,q)\,\Gamma(tz^{-2};p,q)\,
\psi_\lambda(z;p,q,\frac{pq}t)\,.
} As yet another check of consistency of our procedure we verified,
 perturbatively in $p$ and $q$, that the functions
\eigebfuncslist\  indeed satisfy this property.
The norm of the eigenfunctions $\hat\psi_\lambda(z)$ with respect to the vector multiplet measure~\measure\ gives the structure constants $\cal C_\lambda$.
Namely,
\eqn\normconst{
{\cal C}_\lambda = 
\left(\langle\langle 
\hat\psi_\lambda,\;
\hat\psi_\lambda\rangle\rangle
\right)^{\frac12}.
} The norms of the lowest eigenfunctions are given by,
\eqn\norms{
\eqalign{
&\langle\langle 
\hat\psi_0,\;
\hat\psi_0\rangle\rangle=1-2 q^2+(4 p q-6 q^2+p^2 (-6+4 q^2)) T+\dots\,,\cr
&\langle\langle 
\hat\psi_1,\;
\hat\psi_1\rangle\rangle=q-2 q^2+(2 q-6 q^2+p (-2 q+4 q^2)+p^2 (-2 q+4 q^2)) T+\dots\,,\cr
&\langle\langle 
\hat\psi_2,\;
\hat\psi_2\rangle\rangle=q^2+(4 q^2-2 p q^2-4 p^2 q^2) T+\dots\,.
}
}

From here it is immediate to compute the ${\cal I}_1$ index for any $A_1$ quiver. For example,
this index for torus with one puncture is given by

\eqn\torpunc{
\eqalign{
{\cal I}_1^{(1,1)}(a)=\sum_{\lambda=0}^\infty {\hat \psi}_\lambda^{(1)}(a;p,q)=
\left(\frac1a+a\right)q^{1/2}+
4\left(1+\frac1{a^2}+a^2\right)\, q+
2\left(3+\frac1{a^2}+ a^2\right) p q+\cdots\,.
}
} The same index can be computed  by gluing together two legs of the trinion~\trinionAone\ 

\eqn\torusonepunc{
{\cal I}^{(1,1)}_1(a)=
\left[\frac{d}{dT}\kappa_1\oint\frac{dz}{2\pi i z}
\frac{\Gamma(\frac{pq}{t}z^{\pm2};p,q)}
{\Gamma(z^{\pm2};p,q)}\; 
\Gamma(t^{1/2}z^{\pm1}z^{\pm1}\,a^{\pm1};p,q)\right]_{T=0}\,.
} One can  check  expanding in $p$ and $q$ that the two expressions above agree.

\

\newsec{The intergral equations: $A_2$ case}

Let us now discuss the higher rank generalization of the discussion in the previous section.
 Since 
the $A_2$ case is rather generic and we want to be explicit we 
will restrict the discussion to it. 
Unlike the $A_1$ quivers the  higher rank, $A_{N-1}$, theories have many types of different punctures
labeled by partitions of $N$. For example $A_2$ has two non-trivial types
 of punctures, minimal $U(1)$  puncture
($3=2+1$) and maximal $SU(3)$ puncture ($3=1+1+1$).\foot{
The trivial partition $3=3$ corresponds to ``no puncture''.}    
The index of the $A_2$  theory corresponding to Riemann surface with genus $g$, 
$s_1$ $SU(3)$ and $s_2$ $U(1)$ punctures is assumed to be given by~\GaiottoXA,\foot{
As in the $A_1$ case we assume here that the flavor symmetry is large enough. {\it E.g.}, for $g=1$
there is at least one minimal puncture. See~\refs{\GaiottoUQ,\GaddeUV} for a discussion of why and when this expression for the index fails.
}

\eqn\genAtwoind{
{\cal I}^{g,(s_1,s_2)}=
\sum_{\lambda_1\geq \lambda_2=0}^\infty
{\cal C}_{\bf \lambda}^{2g-2+s_1+s_2} \;\prod_{\ell_1=1}^{s_1}\psi_{\bf\lambda}({\bf z}_{\ell_1})\;\prod_{\ell_2=1}^{s_2}\phi_{\bf\lambda}({a}_{\ell_2})\,,
} where $\lambda=(\lambda_1,\lambda_2)$.
The theory  corresponding to the sphere with one minimal and two maximal punctures 
is a free hyper-multiplet in bifundamental representation of $SU(3)\times SU(3)$. The index of this
theory is given by

\eqn\indAtwotrin{
{\cal I}^{0,(2,1)}=\sum_{\lambda_1\geq \lambda_2=0}^\infty
{\cal C}_{\bf \lambda} \;\psi_{\bf\lambda}({\bf z})\;\psi_{\bf\lambda}({\bf y})\;\phi_{\bf\lambda}({a})\;=
\prod_{i,j=1}^N\Gamma\left(t^{\frac12} (z_iy_ja)^{\pm1};\,p,\,q\right)\,.
} 
This is the $A_2$ generalization of \trinioind. Note that we have two un-known functions, 
$\psi_\mu({\bf z})$ and $\phi_\mu(a)$, associated to the $SU(3)$ maximal and the $U(1)$ minimal
punctures respectively. The fugacities ${\bf z}=(z_1,z_2,\frac{1}{z_1z_2})$ label the maximal torus
 of $SU(3)$ 
and $a$ is a $U(1)$ fugacity. The functions $\psi_\mu({\bf z})$ are eigenfunctions of elliptic RS,
but the functions $\phi_\mu(a)$ are not. The latter can be related to certain residues of the
 former~\refs{\GaiottoXA,\GaiottoUQ}.\foot{ This relation won't be too useful for our perturbative technique though.}  The vector multiplet measure under which $\psi_\mu({\bf z})$ are taken to be orthonormal is

\eqn\measureAtwo{
\langle\langle\psi_\mu,\psi_\lambda\rangle\rangle=
\kappa_2\oint\prod_{i=1}^2\frac{dz_i}{2\pi i z_i}
\prod_{i\neq j}
\frac{\Gamma(\frac{pq}{t}z_i/z_j;p,q)}
{\Gamma(z_i/z_j;p,q)}\psi_\mu({\bf z})\psi_\lambda({\bf z}^{-1})=\delta_{\lambda_1\mu_1}\,
\delta_{\lambda_2\mu_2}\,.
}
In principle one can generalize the recursive procedure we described for the $A_1$ case to compute indices of arbitrary $A_{N-1}$ quivers. However, the new ingredient here is that not
for all types of three-punctured spheres the indices are easily computable. For instance, the 
class ${\cal S}$ theories of type $A_2$ have two theories corresponding to two non-trivial three-punctured spheres: one we described 
above \indAtwotrin, and another one with three $SU(3)$, maximal, punctures. The latter theory 
is an interacting strongly-coupled SCFT with no tunable couplings~\MinahanFG. The index for such a theory is hard to compute in general~\GaddeTE. Here knowing explicitly the functions appearing in 
expressions of the form \genAtwoind\ for higher rank theories gives the only known
feasible tool for computing the indices. We thus turn to describe the procedure to determine 
these functions for the $A_2$ case.

\

\noindent From \indAtwotrin\ we can get one integral equation relating 
the functions $\hat \psi_\mu$ and $\hat\phi_\mu$ for the $A_2$ case
\eqn\intAtwoA{
\hat \psi_{\bf\lambda}({\bf y})\;\hat \phi_{\bf\lambda}({a})\;=
\kappa_2\oint\prod_{i=1}^2\frac{dz_i}{2\pi i z_i}
\prod_{i\neq j}
\frac{\Gamma(\frac{pq}{t}z_i/z_j;p,q)}
{\Gamma(z_i/z_j;p,q)}
\prod_{i,j=1}^3\Gamma\left(t^{\frac12} (z_iy_ja)^{\pm1};\,p,\,q\right)\;
\hat \psi_{\bf\lambda}({\bf z}^{-1})\,.
} The hatted eigenfunctions include the structure constants ${\cal C}_\lambda$ in them. This equation
is a straightforward generalization of the $A_1$ equation~\eigenres. 
However, it involves now two unknown functions: one for maximal and one for minimal puncture. 
We need another equation 
to fix both functions. Such an equation is provided by Argyres-Seiberg duality~\ArgyresCN:
two different ways to write the index of the theory associated with a  
sphere with two maximal and two minimal punctures~\GaddeTE. We can write the index of this theory as

\eqn\ASduality{
\eqalign{
&{\cal I}^{0,(2,2)}=\sum_{\lambda_1\geq \lambda_2=0}^\infty
{\cal C}_{\bf \lambda}^{-2} \;\hat\psi_{\bf\lambda}({\bf y_1})\;\hat \psi_{\bf\lambda}({\bf y_2})\;\hat \phi_{\bf\lambda}({a})\;\hat \phi_{\bf\lambda}({b})=\cr
&\sum_{\lambda_1\geq \lambda_2=0}^\infty
{\cal C}_{\bf \lambda}^{-2}
\kappa_1\oint\frac{dz}{2\pi i z}
\frac{\Gamma(\frac{pq}{t}z^{\pm2};p,q)}
{\Gamma(z^{\pm2};p,q)}
\Gamma\left(t^{\frac12} z^{\pm1}s^{\pm1};\,p,\,q\right)\;
\hat \psi_{\bf\lambda}(z\,u,z^{-1}\,u,u^{-2})\,
\hat\psi_{\bf\lambda}({\bf y_1})\;\hat \psi_{\bf\lambda}({\bf y_2})
\,.
}
} Here we have~\refs{\ArgyresCN,\GaddeTE}

\eqn\abus{
s^2=a^3/b^3,\qquad
u^2=1/(a\, b)\,.
}
Using the orthogonality of $\hat\psi_\lambda$s we derive our second integral 
equation

\eqn\intAtwoB{
\hat \phi_{\bf\lambda}(b)\;\hat \phi_{\bf\lambda}({a})\;=
\kappa_1\oint\frac{dz}{2\pi i z}
\frac{\Gamma(\frac{pq}{t}z^{\pm2};p,q)}
{\Gamma(z^{\pm2};p,q)}
\Gamma\left(t^{\frac12} z^{\pm1}s^{\pm1};\,p,\,q\right)\;
\hat \psi_{\bf\lambda}(z\,u,z^{-1}\,u,u^{-2})\,.
} 
Expanding as before in power series in $T$ we will get at each order a system
of two linear equations,coming from~\intAtwoA\ and \intAtwoB, 
in two variables for the two unknown functions.
 The base of the recursion is again the known Schur limit  $T=0$.

\

\subsec{Recursive solution}

The solution of the integral equations \intAtwoA\ and \intAtwoB\ when $t=q$ is known~\refs{\GaddeIK,\GaddeUV}

\eqn\schursolAtwo{
\eqalign{
&\hat\psi^{(0)}_{\mu}({\bf z};q)=
\frac{1}{(1-q)^2(1-q^2)}
\left(
\prod_{i\neq j}^3\prod_{\ell=0}^\infty\frac{1}{1-q^{1+\ell} \, z_i/z_j}
\right)\;
\frac{\chi_\mu({\bf z})}{\chi_\mu(q,q^{-1},1)}\,,\cr
&\hat\phi^{(0)}_{\mu}(a;q)=
\frac{1}{(1-q)(1-q^2)}
\left(
\prod_{\ell=0}^\infty\frac{1}{1-q^{\frac32+\ell} a^{\pm3}}
\right)\;
\frac{\chi_\mu(q^{\frac12}a,q^{-\frac12}a,a^{-2})}{\chi_\mu(q,q^{-1},1)}\,.
}
} Here $\chi_\mu({\bf z})$ are the Schur polynomials for the $A_2$ case. 
The functions $\hat \psi^{(0)}_{\lambda}({\bf z})$ are labeled by irreps of $SU(3)$, $\lambda=(\lambda_1,\,\lambda_2)$ with $\lambda_1\geq\lambda_2$, and they are orthogonal under the vector multiplet measure \measureAtwo\ when $t=q$,

\eqn\schurmeasAtwo{
\kappa_2\prod_{i\neq j}
\frac{\Gamma(\frac{pq}{t}z_i/z_j;p,q)}
{\Gamma(z_i/z_j;p,q)}\;\to\;\Delta_{A_2}({\bf z};q)=
\frac{(q;q)^4}6 \;\prod_{i\neq j}\theta(q\,z_i/z_j;q)\,,\qquad
\qquad
\prod_{i=1}^3z_i=1\,.
}
We define

\eqn\kernelAtwoA{
\eqalign{
&\Delta_{A_2}({\bf z},{\bf y},a;p,q,t)=\cr
&\qquad\kappa_2\, \prod_{i\neq j}^3\frac{\Gamma(\frac{pq}{t}z_i/z_j;p,q)}
{\Gamma(z_{i}/z_j;p,q)}\;
\prod_{i,j=1}^3\Gamma(t^{1/2}(z_iy_j\,a)^{\pm1};p,q)=
\sum_{n=0}^\infty T^n\; \Delta_{A_2}^{(n)}({\bf z},{\bf y},a;p,q)\,,\cr
&\widetilde \Delta_{A_2}(z,a;p,q,t)=\kappa_1\, \frac{\Gamma(\frac{pq}{t}z^{\pm2};p,q)}
{\Gamma(z^{\pm2};p,q)}\;
\Gamma(t^{1/2}z^{\pm1}a^{\pm1};p,q)=
\sum_{n=0}^\infty T^n\; \widetilde \Delta_{A_2}^{(n)}(z,a;p,q)\,.
}}
and

\eqn\pertfuncAtwo{
\eqalign{
&\hat \psi_\lambda(z;p,q,t)=
\sum_{n=0}^\infty T^n\;\hat \psi^{(n)}_\lambda({\bf z};p,q)=
\sum_{n=0}^\infty T^n\;\sum_{\mu=0}^\infty C_{\mu;\;\lambda}^{(n)}(p,q)\;\hat \psi^{(0)}_\mu({\bf z};q)\,,\cr
&\hat \phi_\lambda(a;p,q,t)=
\sum_{n=0}^\infty T^n\;\hat \phi^{(n)}_\lambda(a;p,q)=
\sum_{n=0}^\infty T^n\;\sum_{h=-\infty}^\infty {\widetilde C}_{h;\;\lambda}^{(n)}(p,q)\; a^h\,.
} 
}
The two integral equations of the previous sub-section become,

\eqn\twocofsAtwo{
\eqalign{
&\sum_{\ell=0}^n C_{\mu;\lambda}^{(\ell)}\;
{\widetilde  C}_{h;\lambda}^{(n-\ell)}=
\sum_{\ell=0}^n\sum_{\nu}{\Gamma^{(\ell)}}^{\nu}_{\mu; \, h} 
\; C_{\nu;\lambda}^{(n-\ell)}\,,\cr
&\sum_{\ell=0}^n {\widetilde C}_{g;\lambda}^{(\ell)}\;
{\widetilde  C}_{h;\lambda}^{(n-\ell)}=
\sum_{\ell=0}^n\sum_{\nu}{{\widetilde \Gamma}}^{(\ell)\,\nu}_{g, \, h} 
\; C_{\nu;\lambda}^{(n-\ell)}\,,
}
} where

\eqn\GmmadefAtwoA{
\eqalign{
&{\Gamma^{(k)}}^{\rho}_{h;\,\nu}=
\oint \prod_{i=1}^2\frac{dz_i}{2\pi i z_i}
\oint \prod_{i=1}^2\frac{dy_i}{2\pi i y_i}
\oint \frac{da}{2\pi i a}\cr
&\qquad\qquad
\frac{\Delta_{A_2}^{(k)}({\bf z},{\bf y},a;q,p)\;
\Delta_{A_2}({\bf y};q)\;
\hat \psi^{(0)}_\rho({\bf z}^{-1};q)\hat \psi^{(0)}_\nu({\bf y}^{-1};q)\;a^{-h}\;}
{
\langle \hat \psi^{(0)}_\nu,\, \hat \psi^{(0)}_\nu\rangle}\,,\cr
&{\widetilde \Gamma}^{(k)\rho}_{g,\,h}=
\oint \frac{dz}{2\pi i z}
\oint \frac{da}{2\pi i a}
\oint \frac{db}{2\pi i b}
\;\widetilde \Delta_{A_2}^{(k)}(z,\left(\frac{a}{b}\right)^{\frac32};q,p)\;
\hat \psi^{(0)}_\rho(\frac{z}{\sqrt{ab}},\frac{z^{-1}}{\sqrt{ab}},ab;q)\;a^{-g}\;b^{-h}\,.
}
}

\

\noindent  At the base of the recursion, $n=0$, we have

\eqn\twocofsAtwoBase{
\eqalign{
&\delta_{\mu\lambda}\;
{\widetilde  C}_{h;\lambda}^{(0)}=
{\Gamma^{(0)}}^{\lambda}_{\mu; \, h} 
\,,\qquad\qquad{\widetilde C}_{g;\lambda}^{(0)}\;
{\widetilde  C}_{h;\lambda}^{(0)}=
{{\widetilde \Gamma}}^{(0)\,\lambda}_{g, \, h}\,.
}
} Let us derive the leading correction to the $A_2$ eigenfunctions.  From the first equation in \twocofsAtwo\
we get

\eqn\firsteqAtwo{
C_{\mu;\lambda}^{(1)}\;
\left({\widetilde  C}_{h;\lambda}^{(0)}- {\widetilde  C}_{h;\mu}^{(0)}
\right)+
\delta_{\mu\lambda}\;
{\widetilde  C}_{h;\lambda}^{(1)}=
{\Gamma^{(1)}}^{\lambda}_{\mu; \, h} 
\,,
} and in particular,

\eqn\solsone{
\eqalign{
&\mu\neq\lambda\;:\qquad\qquad
C_{\mu;\lambda}^{(1)}\;
\left({\widetilde  C}_{h;\lambda}^{(0)}- {\widetilde  C}_{h;\mu}^{(0)}
\right)=
{\Gamma^{(1)}}^{\lambda}_{\mu; \, h} \,,\cr
&\mu=\lambda\;:\qquad\qquad
{\widetilde  C}_{h;\lambda}^{(1)}=
{\Gamma^{(1)}}^{\lambda}_{\lambda; \, h} 
\,.
}
} Thus we completely fix ${\widetilde  C}_{h;\lambda}^{(1)}$ and $C_{\mu;\lambda}^{(1)}$
for $\mu\neq \lambda$. To solve for $C_{\lambda;\lambda}^{(1)}$ one needs the second equation
in \twocofsAtwo.
Note also that in the first line in \solsone\ the value of $h$ is arbitrary and thus this equation
encodes a large number of consistency conditions. The second equation in \twocofsAtwo\ for $n=1$ gives

\eqn\secsols{
\eqalign{
& {\widetilde C}_{g;\lambda}^{(1)}\;
{\widetilde  C}_{h;\lambda}^{(0)}+
 {\widetilde C}_{g;\lambda}^{(0)}\;
{\widetilde  C}_{h;\lambda}^{(1)}=
{{\widetilde \Gamma}}^{(1)\,\lambda}_{g, \, h} +
\sum_{\nu}{{\widetilde \Gamma}}^{(0)\,\nu}_{g, \, h} 
\; C_{\nu;\lambda}^{(1)}\,.
}
} From here we deduce

\eqn\secsolF{
C_{\lambda;\lambda}^{(1)}\; {{\widetilde \Gamma}}^{(0)\,\lambda}_{g, \, h}= {\Gamma^{(1)}}^{\lambda}_{\lambda; \, g}\;
{\widetilde  C}_{h;\lambda}^{(0)}+
{\Gamma^{(1)}}^{\lambda}_{\lambda; \, h}
 {\widetilde C}_{g;\lambda}^{(0)}\;
-
{{\widetilde \Gamma}}^{(1)\,\lambda}_{g, \, h} 
-\sum_{\nu\neq \lambda}{{\widetilde \Gamma}}^{(0)\,\nu}_{g, \, h} 
\,\frac{{\Gamma^{(1)}}^{\lambda}_{\nu; \, \hat h}}
{{\widetilde  C}_{\hat h;\lambda}^{(0)}- {\widetilde  C}_{\hat h;\nu}^{(0)}} \,,
} where $\hat h$ is any integer such that ${\widetilde  C}_{\hat h;\lambda}^{(0)}- {\widetilde  C}_{\hat h;\nu}^{(0)}\neq0$.
This completes the determination of the leading corrections and gives yet another set
of consistency equations. For general order in $T$ expansion the first equation in \twocofsAtwo\ with $\mu\neq \lambda$
detemines $C_{\mu;\lambda}^{(n)}$ and with $\mu=\lambda$
it determines $ {\widetilde C}_{h;\lambda}^{(n)}$, while the second equation in \twocofsAtwo\
determines $C_{\lambda;\lambda}^{(n)}$.

\

One can make several consistency checks of the procedure.
Let us mention a simple non-trivial example of such a check.
The flavor symmetry of ${\cal I}^{0,(3,0)}$ is $SU(3)^3$ but is actually supposed to enhance
 to $E_6$~\GaiottoWE.
In particular, the symmetry of ${\cal I}^{1,(1,0)}$ is $G_2$~\GaiottoUQ: we can check that this enhancement of the flavor symmetry indeed occurs for the solution of our perturbative algorithm.
 The index ${\cal I}_1^{1,(1,0)}$ is given by

\eqn\torAtwoB{
\eqalign{
&{\cal I}_1^{(1,(1,0))}({\bf z})=\sum_\lambda \hat\psi^{(1)}_\lambda({\bf z})=
\sum_\lambda \left\{\left(
\frac{{\Gamma^{(1)}}^{\lambda}_{\lambda; \, g}}{
{\widetilde  C}_{g;\lambda}^{(0)}}+
\frac{{\Gamma^{(1)}}^{\lambda}_{\lambda; \, h}}
{ {\widetilde C}_{h;\lambda}^{(0)}}\;
-\frac{{{\widetilde \Gamma}}^{(1)\,\lambda}_{g, \, h} }{{\widetilde  C}_{g;\lambda}^{(0)}{\widetilde  C}_{h;\lambda}^{(0)}}\right)\hat\psi^{(0)}_\lambda({\bf z})+\right.\cr
&\qquad\qquad\qquad\left.\sum_{\nu\neq\lambda}
\left(
\left[\hat\psi^{(0)}_\nu({\bf z})-\hat\psi^{(0)}_\lambda({\bf z})\frac
{{\widetilde  C}_{g;\nu}^{(0)}{\widetilde  C}_{h;\nu}^{(0)}}
{{\widetilde  C}_{g;\lambda}^{(0)}{\widetilde  C}_{h;\lambda}^{(0)}}
\right]\,
\frac{{\Gamma^{(1)}}^{\lambda}_{\nu; \, \hat h}}
{{\widetilde  C}_{\hat h;\lambda}^{(0)}- {\widetilde  C}_{\hat h;\nu}^{(0)}}\right)
\right\}\,.
}
} The fugacities $z_1$ and $z_2$ label the Cartans of $SU(3)$ but actually the symmetry here is
enhanced to $G_2$ and this index has a natural expansion in terms of $G_2$ characters
labeled by the same parameters. To the lowest orders we have

\eqn\Gtwo{
{\cal I}_1^{(1,(1,0))}({\bf z})=2\,q\,(1+p)\,\chi^{G_2}_{14}({\bf z})-
2\,p\,q\,\chi^{G_2}_{7}({\bf z})+\cdots\,.
} This expression can be checked against the results of~\GaddeTE\ for the index ${\cal I}^{0,(3,0)}$
 obtained by using an inversion formula~\SpirWarnaar\ of the integral equation at hand~\intAtwoB.

Yet another, simpler, consistency check of our procedure is the computation
of  ${\cal I}_1^{1,(0,1)}$ which is given by

\eqn\torAtwoA{
{\cal I}_1^{(1,(0,1))}(a)=\sum_\lambda \hat\phi^{(1)}_\lambda(a)=
\sum_\lambda \sum_{h=-\infty}^\infty {\Gamma^{(1)}}^{\lambda}_{\lambda; \, h} \;a^h\,,
} and can be explcitly seen to match, expanding in $p$ and $q$, the same index computed
by gauging a diagonal combination of two $SU(3)$ symmetries in ${\cal I}^{0,(2,1)}$ given in~\indAtwotrin.

\

\newsec{Comments}

Let us make several brief remarks about our results and further research directions.

\item 1.  Assuming S-duality and making certain mathematical assumptions about the structure of the 
index of theories of class ${\cal S}$ we derived a simple perturbative algorithm 
to compute the index. This algorithm involves determining a set of eigenfunctions
of the elliptic RS model. We gave explicit details of the $A_1$ and the $A_2$ cases.
The knowledge of the eigenfunctions up to a finite order in the expansion parameter 
is sufficient to compute exactly a (finite) set of indices generalizing the Schur index.

\item 2. The system of recursive equations determining the eigenfunctions needed to compute the index is highly overdetermined.  We have checked that our  procedure is consistent to lowest orders in the expansion parameter.  It would be very beneficial to verify/prove whether the procedure makes sense to all orders. In particular it should be enlightening in many respects to obtain closed form expressions for the relevant eigenfunctions.

\item 3. The algorithm discussed in this note, although stated for $A_1$ and $A_2$ cases, can be generalized in principle to higher rank theories. There we have more types of punctures with 
a priori un-known functions associated to each one of them.
But we also have more non-trivial dualities~\GaiottoWE\ generalizing 
the Argyres-Seiberg duality which can be translated into integral equations relating the different un-known functions. It should be interesting to study the general 
structure of these systems of integral equations.

\item 4. As a base of our recursive procedure we have
 used the known expressions for the index in the Schur case, $t=q$. 
One can try and use the known solution of the integral equations in the Macdonald case,
$p=0$, as a starting point for the perturbative construction of the eigenfunctions.

\

\

\bigskip
\noindent{\bf Acknowledgments:} 

\

We would like to thank  
A.~Gadde,  D.~Gaiotto, E.~Langmann, L.~Rastelli, Y.~Tachikawa, and B.~Willett for very interesting discussions.
We are grateful  to  the organizers of the workshop ``Elliptic Integrable Systems and Hypergeometric Functions '', (July 2013, Lorentz Center), the organizers of the Simons Summer Workshop in Stony Brook (August 2013),
and to the high energy theory group at the Weizmann institute for hospitality during different stages of this project.We gratefully acknowledge support from the Martin~A.~Chooljian and Helen Chooljian membership
 at the Institute for Advanced Study. This research was also partially supported by
NSF grant number PHY-0969448.

\listrefs
\end